\documentclass[a4paper,fleqn,usenatbib]{mnras}

\usepackage{newtxtext,newtxmath}

\usepackage[T1]{fontenc}
\usepackage{ae,aecompl}

\usepackage{graphicx}	
\usepackage{amsmath}	
\usepackage{amssymb}	

\usepackage{times}
\usepackage{color}
\usepackage{subfigure}

\usepackage{multirow}
\usepackage{hyperref}
\def\apjl{ApJ}%
\def\apjs{ApJS}%
\def\ao{Appl.~Opt.}%
\def\aap{A\&A}%
\title[The {\em Herschel}-ATLAS Data Release 2 Paper III]{The second {\em Herschel}\thanks{{\em Herschel}
is an ESA space observatory with science instruments provided
by European-led Principal Investigator consortia and with important participation from NASA.}-ATLAS Data Release - III: optical and near-infrared counterparts in the North Galactic Plane field}

\author[C. Furlanetto et al.]{C. Furlanetto$^{1,2}$\thanks{E-mail: cristina.furlanetto@ufrgs.br},
S. Dye$^{2}$, N. Bourne$^{3}$, S. Maddox$^{3,4}$, L. Dunne$^{3,4}$, S. Eales$^{4}$,
\newauthor
E. Valiante$^{4}$, M. W. Smith$^{4}$, D. J. B. Smith$^{5}$, R. J. Ivison$^{3,6}$, E. Ibar$^{7}$
\\
$^{1}$Instituto de F\'isica, Universidade Federal do Rio Grande do Sul, Av. Bento Gon\c{c}alves, 9500, 91501-970, Porto Alegre, RS, Brazil\\
$^{2}$School of Physics and Astronomy, Nottingham University, University Park, Nottingham, NG7 2RD, UK\\
$^{3}$Institute for Astronomy, University of Edinburgh, Royal Observatory, Edinburgh EH9 3HJ, UK\\
$^{4}$School of Physics and Astronomy, Cardiff University, The Parade, Cardiff, CF24 3AA, UK\\
$^{5}$Centre for Astrophysics Research, School of Physics, Astronomy and Mathematics, University of Hertfordshire, College Lane, Hatfield, AL10 9AB, UK\\
$^{6}$European Southern Observatory, Karl Schwarzschild Strasse 2, D-85748 Garching, Germany\\
$^{7}$Instituto de F\'isica y Astronom\'ia, Universidad de Valpara\'iso, Avda. Gran Breta\~na 1111, Valpara\'iso, Chile
}

\date{Accepted XXX. Received YYY; in original form ZZZ}

\pubyear{2017}

\begin{document}

\defcitealias{bourne16}{B16}
\defcitealias{Fleuren2012}{F12}
\defcitealias{valiante16}{V16}
\defcitealias{Smith2011}{S11}

\label{firstpage}
\pagerange{\pageref{firstpage}--\pageref{lastpage}}
\maketitle 
 
\begin{abstract}
This paper forms part of the second major public data release of the {\em Herschel} Astrophysical Terahertz Large Area Survey (H-ATLAS). In this work, we describe the identification of optical and near-infrared counterparts to the submillimetre detected sources in the $177$ deg$^2$
North Galactic Plane (NGP) field. We used the likelihood ratio method to identify counterparts in 
the Sloan Digital Sky Survey and in the UKIRT Imaging Deep Sky Survey within a search radius of 
$10$ arcsec of the H-ATLAS sources with a $4\sigma$ detection at $250 \, \mu$m. We obtained reliable 
($R \ge 0.8 $) optical counterparts with $r< 22.4$ for 42429 H-ATLAS sources ($37.8$ per cent), with an estimated completeness 
of $71.7$ per cent and a false identification rate of $4.7$ per cent. 
We also identified counterparts in the near-infrared using deeper
$K$-band data which covers a smaller $\sim25$\,deg$^2$. We found reliable near-infrared 
counterparts to $61.8$ per cent of the $250$-$\mu$m-selected sources within that area. 
We assessed the performance of the likelihood ratio method to identify optical 
and near-infrared counterparts taking into account the depth and area of 
both input catalogues. Using catalogues with the same 
surface density of objects in the overlapping $\sim25$\,deg$^2$ area, we obtained that 
the reliable fraction in the near-infrared ($54.8$ per cent) is significantly higher 
than in the optical ($36.4$ per cent). Finally, using deep radio data which covers a small region of the NGP field, we found that $80 - 90$ per cent of our reliable identifications are correct.

\end{abstract}




\section{Introduction}

This paper presents the identification of counterparts to detected submillimetre (submm)
sources for the second major public data release of the
{\em Herschel} Astrophysical Terahertz Large Area Survey (H-ATLAS). H-ATLAS is the largest 
single key-project in area carried out in open
time with the Herschel Space Observatory \citep{Pilbratt2010}. 
In its entirety, H-ATLAS covers approximately 600\,deg$^2$ in five photometric 
bands: 100, 160, 250, 350, 500 \,$\mu$m. 
The area is split into three distinct regions, selected to avoid bright continuum emission from 
dust in the Galaxy and to maximize the amount of data in other wavebands:  
the north galactic plane (NGP) comprising a single contiguous 177\,deg$^2$ field 
centred at approximately (200$^\circ$,29$^\circ$); three equatorial fields which
cover a total area of 161\,deg$^2$ and coincide with the
equatorial areas surveyed in the Galaxy And Mass Assembly (GAMA)
redshift survey \citep{driver11,liske15}; and a strip in the vicinity of the
southern galactic plane (SGP) with an area of $\sim$317.6\,deg$^2$. 
Full details of the survey design are given in \citet{eales10}. Note that the 
survey geometry reported for the SGP field in
this reference has since been superseded with a single contiguous strip.

The first public data release (DR1) of H-ATLAS covered the equatorial fields. 
Details of the processing and characterisation
of image data and submm source catalogues within DR1 were given in
\citet[][V16 hereafter]{valiante16}. Optically identified counterparts
to the submm detected H-ATLAS sources in DR1 were discussed by
\citet[][B16 hereafter]{bourne16}. The DR1 release also includes the area previously known as the Science Demonstration Phase (SDP), which was described in \citep{ibar10}, \citep{rigby11}, \citep{pascale11} and \citep{Smith2011}.

The second public data release (DR2) products encompass the NGP and SGP fields. The DR2 is described herein and in two accompanying papers, \citet[][Paper I]{PaperI} and \citet[][Paper II]{PaperII}. Paper I describes the {\em Herschel} images of the NGP and SGP fields and an investigation of their noise properties. Paper II presents the catalogues of submm sources detected on the images. The imaging 
and source catalogue products from the NGP and SGP fields can be
obtained from the H-ATLAS web page\footnote{\tt http://www.h-atlas.org/}. 

In this paper, we describe the identification of submm source
counterparts across the whole NGP field in the optical via the Sloan
Digital Sky Survey \citep{abazajian09} and their corresponding matches
in the near-infrared (near-IR) via the UKIRT Infrared Deep Sky Survey
\citep[UKIDSS;][]{lawrence07}.  In addition, we investigate the
identification of counterparts in the $K$ band within a deeper
$\sim 25$\,deg$^2$ subset of the NGP field observed with UKIRT, with a view to
understanding the properties of sources not identified in the
shallower optical and near-IR data. The imaging and catalogue products 
of this deeper survey in the $K$ band are also part of the H-ATLAS DR2. Finally, we carry out an
assessment of the performance of the likelihood ratio technique used
in identifying counterparts by comparing to deep radio interferometric
data which covers a small region of the NGP, common to both the
optical and deeper $K$ band datasets.

The layout of this paper is as follows. In sections \ref{sec_id_optical} and \ref{sec_id_ir}
we detail identification of counterparts to the submm detected sources
in the optical and near-infrared respectively. In Section \ref{comparison} we compare 
the performance of the likelihood ratio method in identifying counterparts 
in the optical and near-infrared. Section \ref{sec_id_radio}
investigates submm source counterparts detected in radio
interferometric data. Finally, we summarize our analysis and the data
release products in section \ref{sec_conclusions}.

\section{Optical counterparts to submm sources}
\label{sec_id_optical}

In this section we present the optical identifications of the 250 $\mu$m SPIRE sources. We use the Sloan Digital Sky Survey because of its astrometric accuracy and complete coverage of the NGP field.

The low angular resolution of {\it Herschel} observations and the intrinsic faintness of the counterpart due to dust obscuration are limiting factors when characterizing the submm sources at longer wavelengths. The large positional uncertainties of bright submm sources and the presence of multiple possible counterparts within the large beam (due to the high surface density of objects in optical surveys) means that the identification of counterparts must rely on statistical methods (e.g. \citealt{Smith2017}). 

One method often applied to decide which objects are truly associated and which are unrelated background or foreground objects is the likelihood ratio (LR) method \citep{Sutherland, Ciliegi03}. The LR method was used to identify optical and near-infrared counterparts to SPIRE sources in previous releases of H-ATLAS data. For example, the technique was adopted by \citet[S11 hereafter]{Smith2011} to identify SDSS counterparts in the H-ATLAS Science Demonstration Phase, by \citet{bond12} to identify the Wide-field Infrared Survey (WISE) counterparts to the sources of Phase-1 GAMA15 field.
, by \citet[F12 hereafter]{Fleuren2012} to identify VISTA Kilo-degree Infrared Galaxy Survey (VIKING) counterparts in the Phase-1 GAMA9 field, and by \citetalias{bourne16} to identify SDSS counterparts in the three equatorial H-ATLAS fields in the DR1.

In the following we describe the application of the LR method to identify optical counterparts to H-ATLAS NGP sources. In order to consistently identify the optical counterparts across all H-ATLAS fields, we adopted the same assumptions as in the analysis of \citetalias{bourne16} for DR1.

\subsection{Optical data}
\label{optical_data}

We constructed our optical object catalogue by selecting all primary objects (the ``main'' observations given multiple observations of the objects) in the 10th data release \citep[DR10;][]{ahn14} of SDSS with $r_{\rm model}< 22.4$ in the NGP field. The $r_{\rm model}= 22.4$ limit is essentially the completeness limit of SDSS photometric survey.  There are 2744529 objects satisfying these criteria included in our optical object catalogue. Unlike \citetalias{Smith2011} and \citetalias{bourne16}, which used SDSS DR7 to identify optical counterpart to H-ATLAS sources, we opted for DR10 because it contains a larger sample of objects with spectroscopic redshifts when compared to previous releases.

In order to remove spurious objects from the catalogue, usually associated with erroneous deblends of nearby galaxies or diffraction spikes of stars, we visually inspected all SDSS objects with deblend flags within 10 arcsec of each SPIRE source. In this process, we removed 3671 SDSS objects from the optical input catalogue. 

Since the H-ATLAS NGP field has almost complete near-infrared (NIR) coverage from the UKIDSS Large Area Survey \citep{lawrence07}, we added the YJHK photometry from its 9th data release to our optical object catalogue by performing a simple nearest neighbour matching to the SDSS objects. We found that $61.6$ per cent of the primary SDSS DR10 candidates in NGP field have a match in UKIDSS-LAS within 3 arcsec. By generating a catalogue of random positions with the same source density as UKIDSS-LAS and matching it to our optical object catalogue, we estimated that the probability of false UKIDSS-LAS association is smaller than 3 per cent. The remaining SDSS candidates are too faint to be detected in the UKIDSS-LAS. 

Our optical object catalogue contains 39073 spectroscopic redshifts ($z_{\rm spec}$) obtained with SDSS DR10,  corresponding to $1.4$ per cent of the sample. In addition, a further 1128 spectroscopic redshifts were added from the CfA Redshift Survey \citep{cfa}. Given the lack of a quality flag in the latter redshift catalogue, for objects that have redshifts in both surveys, we use the CfA redshift only if there is a redshift warning flag for the object in SDSS DR10. 

For the objects without spectra, photometric redshifts were obtained from SDSS DR10 (for details see \citealp{Csabai2007}).

\subsubsection{Star-galaxy separation}
\label{sg_sdss}

\begin{figure}
\includegraphics[scale=0.32]{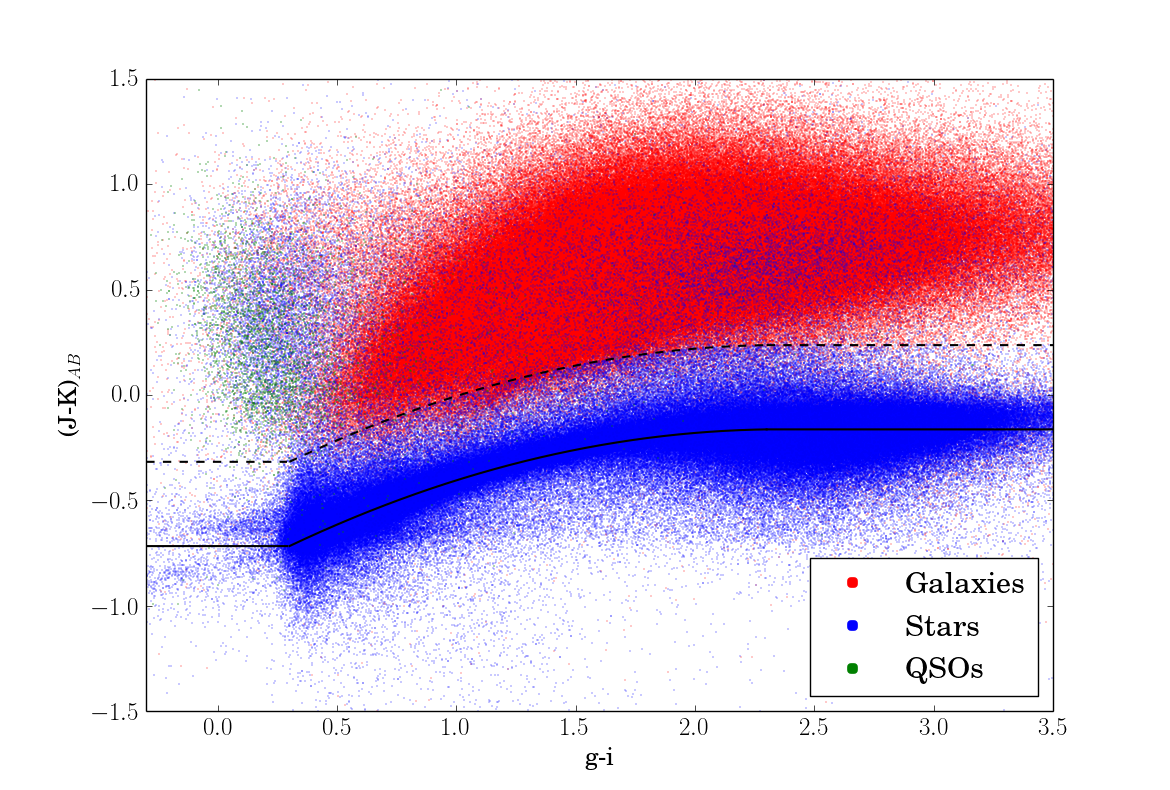}
\caption{The colour-colour diagram of SDSS objects with UKIDSS-LAS identifications. We used a slightly modified version of the relationship of Baldry et al. (2010) 
to separate stars (blue points), galaxies (red points) and QSOs (green points). The solid line shows the stellar locus, obtained from fitting a quadratic
equation (see equation \ref{stellar_locus}) to the combined SDSS/UKIDSS-LAS data set. The dashed line is offset +0.4mag from the stellar locus, representing one of the adopted separation criteria between stars and galaxies. Objects without UKIDSS-LAS counterpart are classified according to their $r_{\rm model}$ magnitude only. }
\label{sgsep}
\end{figure}

We separated stars and galaxy populations in our optical object catalogue following a similar prescription to that used in \citetalias{Smith2011} and \citetalias{bourne16}. This consists of a slightly modified version of the procedure described in \citet{Baldry2010} to select the galaxy sample for the GAMA input catalogue by using a combination of shape and colour parameters. The colour criteria are based on our SDSS-UKIDSS-matched catalogue, as  illustrated in the colour-colour diagram of Fig. \ref{sgsep}. In summary, an object is classified as a galaxy if it satisfies the following constraints
\begin{align}
&\Delta_{\rm sg} > 0.25 \nonumber \\
&\mbox{or}\nonumber \\
&\Delta_{\rm sg} > 0.05 \mbox{~~and~~} \Delta_{\rm sg,jk}>0.40,
\label{sg_criteria}
\end{align}
where $\Delta_{\rm sg}$ is the SDSS star-galaxy separation parameter that quantifies the fraction of extended flux, defined as 
\begin{equation}
 \Delta_{\rm sg} = r_{\rm psf} - r_{\rm model},
 \label{delta_sg}
\end{equation}
where $r_{\rm psf}$ is the r-band magnitude determined from a fit using the point spread function, and
\begin{equation}
 \Delta_{\rm sg,jk} = J_{\rm AB} - K_{\rm AB} - f_{\rm locus}(g-i)
 \label{delta_sgjk}
\end{equation}
is a new star-galaxy separation parameter defined as the $J-K$ separation from the stellar locus, $f_{\rm locus}$, in the $J-K$ versus $g-i$ colour space (see Fig. \ref{sgsep}). The stellar locus in this colour space is defined as in \citet{Baldry2010}
\begin{equation}
\label{stellar_locus}
f_{\rm locus}(x) = \begin{cases}
-0.7172, &  x< 0.3\\
-0.89 + 0.615x-0.13x^2, &  0.3< x <2.3\\
-0.1632,  & x> 2.3, \\
\end{cases}
\end{equation}
where $x=g-i$. The stellar locus is shown as the black solid line in colour-colour diagram of Fig. \ref{sgsep}. In this process we used model magnitudes $g$ and $i$ from SDSS DR10 and $J$ 
and $K$ magnitudes (2-arcsec aperture) from UKIDSS-LAS. 

For the remaining SDSS objects in our optical object catalogue not satisfying the first part of equation (\ref{sg_criteria}) and with no UKIDSS counterpart, galaxies are defined as objects satisfying 
\begin{equation}
\Delta_{\rm sg} > f_{\rm sg,slope}(r_{\rm model}) \mbox{~~~~~(no $J-K$ measurement)},
\label{sgsep_nomatch}
\end{equation}
where 
\begin{equation}
f_{\rm sg,slope}(x) = 
\begin{cases}
0.25, &  x< 19.0\\
0.25 -\frac{1}{15}(x-19), &  19.0< x <20.5\\
0.15,  & x> 20.5. \\
\end{cases}
\end{equation}

We also added the constraint $z_{\rm spec}>0.001$ to the classification above for the objects with spectroscopic redshift available.

We identified the objects that do not satisfy any of the constraints above as unresolved. Among this class of objects, those 
with $z_{\rm spec}>0.001$ were classified as quasars (QSOs). The remaining unresolved objects were classified as stars. 

Fig. \ref{sgsep} shows the colour-colour diagram of the objects in our optical object catalogue with a near-infrared counterpart, 
the location of the stellar locus and the separation criteria used. We identified 1981642 galaxies (72.30 per cent of the sample), 754999 stars (27.55 per cent) and 4216 quasars (0.15 per cent). There are 3439 objects classified as quasars with $z_{\rm spec}>1$. Among the stellar sample, there are 171018 unresolved sources whose classification was based purely on the optical information. Those correspond to $16.2$ per cent of the optical sources that do not have a near-infrared counterpart in UKIDSS-LAS (i.e. are not in Fig. \ref{sgsep}) and to $6.2$ per cent of the entire sample. It is important to note that this stellar subsample is more likely to be biased by faint unresolved galaxies.

\subsection{Likelihood ratio analysis}
\label{SecLR}

We used the likelihood ratio method to identify the most reliable optical counterparts to the 250-$\mu$m sources selected in the H-ATLAS NGP field. 
The likelihood ratio method was developed by \citet{Sutherland} and received additional
improvements by \citet{Ciliegi03}, \citet{brusa07} and \citet{chapin11}.

The likelihood ratio technique uses the ratio between the probability of a match being the correct identification and the corresponding probability of being an unrelated background object. The method uses the intrinsic positional uncertainty of the sources and the magnitude distributions of the true counterparts as well as of the background objects. 

The likelihood ratio is defined as
\begin{equation}
 L = \frac{q(m,c)f(r)}{n(m,c)},
 \label{LR}
\end{equation}
where $q(m,c)$ is the probability distribution of the true counterparts with magnitude $m$ and class $c$ (e.g. star/galaxy or other additional property), $f(r)$ is the probability distribution of the source positional errors, and $n(m,c)$ is the magnitude distribution of the unrelated background objects with class $c$. 

The reliability $R_j$ is the probability that an object $j$ is the correct identification of a given SPIRE source. It is defined as
\begin{equation}
R_j= \frac{L_j}{\sum_i L_i + (1-Q_0)},
 \label{reliability}
\end{equation}
where the sum in the denominator is to account for the presence of other potential counterparts to the same SPIRE source and the term $(1-Q_0)$ is the probability that there is no counterpart in the optical survey (see Section \ref{n_q}). We consider objects with $R_j \ge 0.8$ as reliably identified counterparts to the SPIRE source, following \citetalias{Smith2011} and \citetalias{bourne16}.

In the following we describe how the quantities described above are measured.

\subsubsection{Estimation of $f(r)$}
\label{f_r_sec}

As probability distribution of positional errors we adopted a Gaussian distribution with standard deviation $\sigma_{\rm pos}$:
\begin{equation}
 f(r)=\frac{1}{2\pi\sigma^2_{\rm pos}}{\rm exp}\left(\frac{-r^2}{2\sigma^2_{\rm pos}}\right),
 \label{f_r}
\end{equation}
where $r$ is the offset between the $250~\mu$m and $r$-band positions and $\sigma_{\rm pos}=\sqrt{\sigma_{\alpha} \sigma_{\delta}}$ is the geometric mean of positional errors in RA and Dec of the submillimetre source with respect to the optical position.
This expression is based on the assumption that the sources extracted in H-ATLAS 250-$\mu$m maps are point-like sources. 

The positional error $\sigma_{\rm pos}$ is empirically estimated using the offset distribution of all potential counterparts. Following the methods of \citetalias{Smith2011} and \citetalias{bourne16}, we derived a two-dimensional histogram of the separation in RA and Dec of the SDSS objects within a 50-arcsec box around each SPIRE source. We modelled this distribution as consisting of three components: the contribution of the background density, which is constant across the histogram; the contribution from real counterparts and the contribution from other correlated sources due to the clustering of the SDSS objects. We can describe these components by the equation:
\begin{equation}
n(\Delta {\rm RA},\Delta {\rm Dec})=n_0 +Q_0 f(r)+w(r)*f(r),
\label{histogram_distribution}
\end{equation}
where $n_0$ is the constant background density of our optical object catalogue and $Q_0$ is the fraction of the true counterparts that are detected in SDSS. The additional contribution of the nearby SDSS objects that are correlated with the SPIRE source due to the galaxy clustering (but are not the correct counterparts) is given by the angular cross-correlation function $w(r)$ between SPIRE and SDSS positions convolved with the positional error function $f(r)$. The cross-correlation between SPIRE and SDSS samples is modelled as a power-law
\begin{equation}
w(r)=\left( \frac{r}{r_0} \right)^{\delta},
\label{wr}
\end{equation}
where $r_0$ is the correlation length and $\delta$ is the power-law index. 

We assumed that the SDSS positional errors are negligible in comparison to the SPIRE errors, so the width ($\sigma_{\rm pos}$) of $f(r)$ is simply the SPIRE positional error, which can be obtained in $\Delta$RA and $\Delta$Dec from the modelling of the two-dimensional positional offset histogram. 

According to \citet{Ivison2007}, the theoretical form for $\sigma_{\rm pos}$ depends on the full-width at half-maximum (FWHM) and the signal-to-noise ratio (SNR) of the 250$\mu$m detection:
\begin{equation}
 \sigma_{\rm th}(SNR)=0.6\frac{{\rm FWHM}}{{\rm SNR}}.
 \label{sigma_pos_the}
\end{equation}

\citet{bourne14} showed that redder and brighter submm sources have optical associations with a broader distribution of positional offsets than would be expected if these offsets were due to random positional errors in the source extraction. They concluded that this effect is most likely to be explained by the significant contribution from foreground structures in the line of sight to the SPIRE sources, which are not physically associated but may be lensing the source. This interpretation is supported by the lens modelling of the H-ATLAS sample in \citet{gonzalez17}. In order to avoid this bias, which can increase the $L$ values for optical associations to red SPIRE sources, we measured the $\sigma_{\rm pos}$ from the offset histogram of blue SPIRE sources with $S_{\rm 250}/S_{\rm 350} > 2.4$, following the method described in \citetalias{bourne16}. 

\begin{figure}
\includegraphics[scale=0.30]{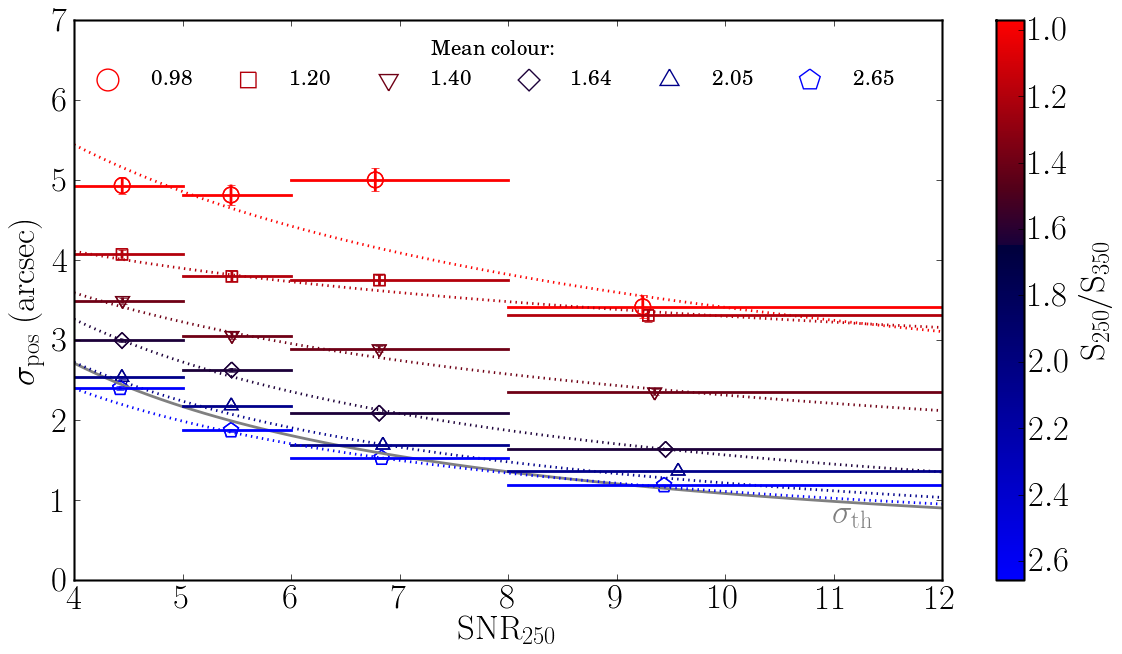}
\caption{$\sigma_{\rm pos}$ measured from the fitting of the two-dimensional offset histogram as a function of the mean value of $250~\mu$m SNR in each SNR bin for six bins of $S_{\rm 250}/S_{\rm 350}$ colour. The symbols indicate the colour bins. The grey solid line corresponds to the theoretical prediction for $\sigma_{\rm pos}$ given in equation (\ref{sigma_pos_the}). Dotted lines show the best-fitting models for $\sigma_{\rm pos}$ as a function of SNR for each colour bin. The results for bluest colour bin are summarized in Table \ref{best-fit-sigmapos}. }
\label{sigma_pos_cbins}
\end{figure}

We fitted the two-dimensional histogram of the separation in RA and Dec using the model given in equation (\ref{histogram_distribution}) to obtain the width $\sigma_{pos}$ of the Gaussian positional errors $f(r)$, the background density $n_0$ and fraction of the true counterparts that are detected in SDSS $Q_0$. During the fitting process, we fixed the power-law parameters of $w(r)$ to the values obtained by \citet{bourne14} for the angular cross-correlation function between SPIRE and SDSS sources ($\delta=-0.7$ and $r_0$ as given in Table \ref{best-fit-sigmapos}). We examined the behavior of $\sigma_{pos}$ as a function of the SNR. In Table \ref{best-fit-sigmapos} we present the best-fitting parameters obtained in the modelling of SDSS positional offsets to blue SPIRE sources in four bins of $250~\mu$m SNR. The dependence of the width $\sigma_{pos}$ as a function of SNR is shown in Fig. \ref{sigma_pos_cbins}. In the same figure we also show the colour dependence of $\sigma_{pos}$ in six bins of colour to illustrate the need to measure this width using blue SPIRE sources due to the bias discussed above, which is likely to be caused by lensing. The fitting results summarized in Table \ref{best-fit-sigmapos} correspond to the bluest colour bin in Fig. \ref{sigma_pos_cbins}.

\begin{table}
\begin{center}
\begin{tabular}{ c c c c c }
\hline
${\rm SNR}_{\rm 250}$ & $N_{\rm SPIRE}$ & $r_{\rm 0}$ (arcsec) & $\sigma_{\rm pos}$ (arcsec) & $Q_0$ \\
\hline
4 - 5 & 3285 &  0.20 $\pm$ 0.02  & 2.40 $\pm$ 0.02 &  0.699 $\pm$ 0.008\\
5 - 6 & 1583 &  0.61 $\pm$ 0.05  & 1.88 $\pm$ 0.02 &  0.810 $\pm$ 0.009\\
6 - 8 & 1316 &  0.38 $\pm$ 0.05  & 1.53 $\pm$ 0.01 &  0.873 $\pm$ 0.008\\
8 - 12 & 799 &  0.38 $\pm$ 0.08  & 1.19 $\pm$ 0.01 &  0.924 $\pm$ 0.008\\
\hline
\end{tabular}
\end{center}
\caption{Results of the fitting of positional offsets between SDSS and blue SPIRE sources with $S_{\rm 250}/S_{\rm 350}>2.4$ in bins of $250~\mu$m SNR. The values for the correlation length $r_0$ were taken from \citetalias{bourne16}. }
\label{best-fit-sigmapos}
\end{table}

We modelled the dependence of the positional errors on the SNR as a power-law:

\begin{equation}
 \sigma_{\rm pos}(SNR)=\sigma_{\rm pos}(5)\left( \frac{SNR}{5} \right)^{\alpha}.
 \label{sigma_pos_snr}
\end{equation}

The best-fitting model for each colour bin is shown as dotted lines in Fig. \ref{sigma_pos_cbins}. For the bluest bin (SPIRE sources with $S_{\rm 250}/S_{\rm 350}>2.4$), we obtained that $\alpha=-0.84\pm0.07$ and $\sigma_{\rm pos}(5)=1.99\pm0.05$ arcsec. This result shows that the empirical dependence of $\sigma_{\rm pos}$ on SNR is not significantly different from the theoretical prediction of equation (\ref{sigma_pos_the}).

We adopted the description of equation (\ref{sigma_pos_snr}) for the expected positional errors used to compute $f(r)$ in our likelihood ratio calculations. In order to avoid unrealistically small errors for very bright sources, we imposed a restriction that $\sigma_{\rm pos} > 1$ arcsec, as in \citetalias{Smith2011}. This minimum positional error also accounts for the possibility that the submm and optical emission may not arise at exactly the same position in a galaxy.

\subsubsection{Estimation of $n(m)$ and $q(m)$}
\label{n_q}

We computed the probability distributions $n(m)$ and $q(m)$ for extragalactic objects (galaxies and QSOs) and for stars separately. 

The $n(m)$ term in equation (\ref{LR}) corresponds to the probability density that a given SDSS source has magnitude $m$. It is estimated from the object counts of the optical object catalogue normalized to the area. 

\begin{figure}
\includegraphics[scale=0.35]{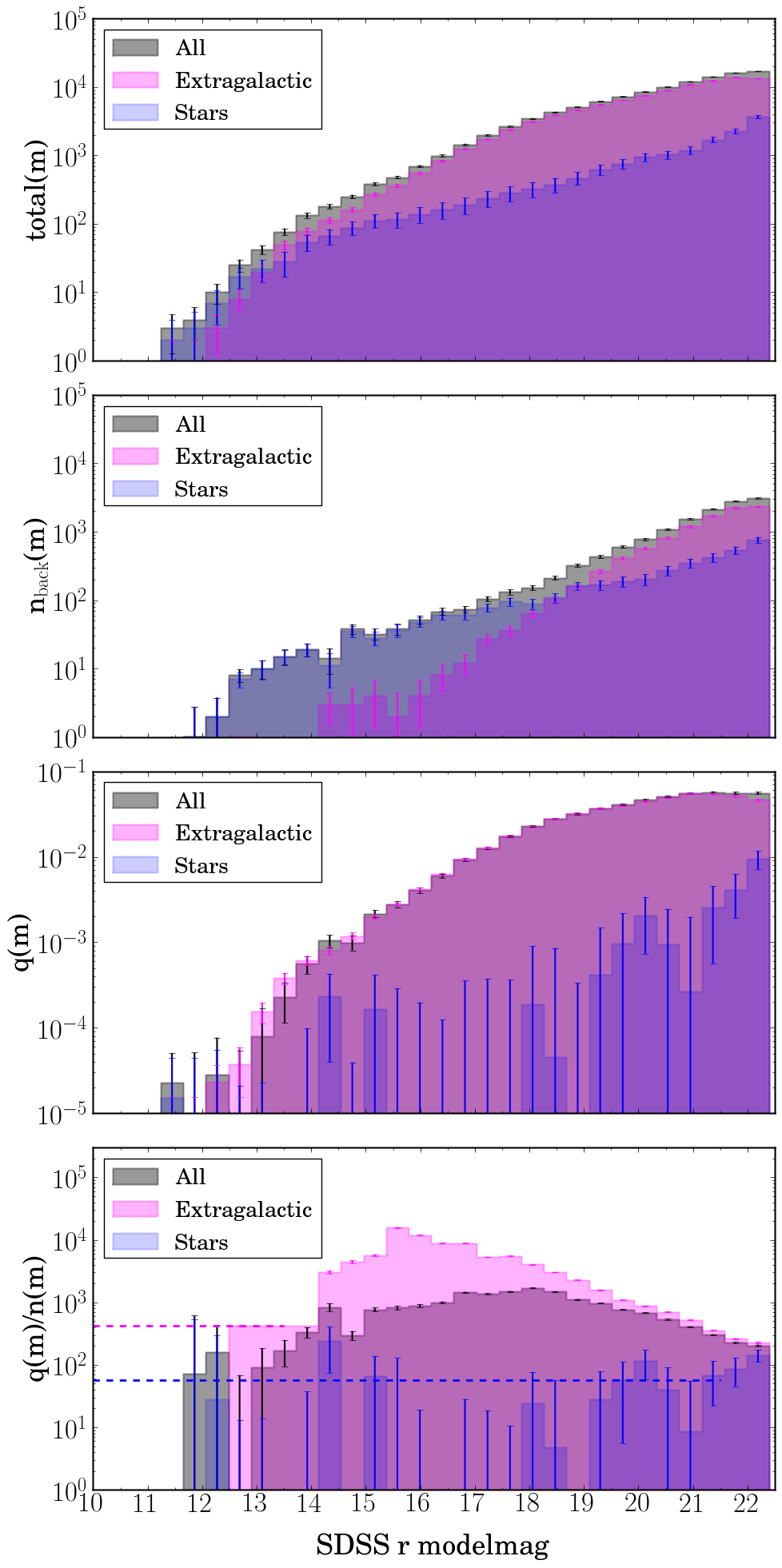}
\caption{The $r$-band magnitude distributions measured by the likelihood ratio method. Each of these distributions is divided into extragalactic objects and stars. The upper panels show the magnitude distribution of all (total) and background ($n_{\rm back}$) SDSS objects within 10 arcsec of the SPIRE sources, respectively. The magnitude distribution of the true counterparts, real(m), is given by subtracting $n_{\rm back}(m)$ from ${\rm total}(m)$. The $q(m)$ distribution describes the probability density that the true optical counterpart has magnitude $m$. The lower panel shows the ratio of the magnitude distributions of true counterparts and of the background objects that is used to compute the likelihood ratio. The $n(m)$ distribution is obtained by normalizing  $n_{\rm back}(m)$. We assume a constant value corresponding to the average of $q(m)/n(m)$ at magnitudes $m<21.5$ for stars and $m<13.5$ for galaxies, since at brighter magnitudes the distributions are not well sampled.}
\label{mag_dist}
\end{figure}

The distribution $q(m)$ is the probability that a true counterpart to a SPIRE source has magnitude $m$. This distribution is estimated using the method described in \citet{Ciliegi03}, which begins by counting all objects in the optical catalogue with magnitude $m$ and within a fixed search radius $r_{\rm max}$ around each SPIRE source to give ${\rm total}(m)$. The contribution of the background source counts, $n_{\rm back}(m)=n(m) N \pi r_{\rm max}^2$, is subtracted from this distribution, producing the magnitude distribution of all true counterparts, ${\rm real}(m)$, which is given by
\begin{equation}
{\rm real}(m)={\rm total}(m)-n_{\rm back}(m),
\label{real_m}
\end{equation}
where $N$ is the number of SPIRE sources. In our analysis, we used $r_{\rm max}=10$ arcsec. The $q(m)$ distribution is then derived by normalizing ${\rm real}(m)$ and scaling it by the overall probability $Q_0$
\begin{equation}
 q(m)=Q_{\rm 0}\frac{\mbox{real}(m)}{\sum_m \mbox{real}(m)}.
 \label{q_m}
\end{equation}
The term $Q_{\rm 0}$ is an estimate of $Q$ 
\begin{equation}
Q=\int_{0}^{m_{\rm lim}} q(m)dm,
\end{equation}
which is the fraction of all true counterparts that are above the SDSS magnitude limit $M_{\rm lim}$. The distributions ${\rm total}(m)$, $n_{\rm back}(m)$ and $q(m)$ derived by this method are shown in Fig. \ref{mag_dist}.

We estimated the normalization term $Q_0$ for stars and extragalactic objects following the method outlined in \citetalias{Fleuren2012}. To avoid multicounting counterparts due to 
clustering or genuine multiple counterparts, which can overestimate the value of $Q_0$, this method measures $1-Q_{\rm 0}$ by counting objects without any counterpart candidate within the 
search radius. These objects are referred to as ``blanks''. The fraction of SPIRE sources that are true blanks is given by the ratio between the observed number of SPIRE blanks and the number of 
random blanks for a given search radius. An estimate of $Q_{\rm 0}$ that is independent of the radius can be obtained by computing $1-Q_{\rm 0}$ for radii in the range 1 to 15 arcsec and 
modelling the dependence of the true blanks on the search radius as
\begin{equation}
B(r) = 1-Q_{\rm 0}F(r),
\label{q0_model}
\end{equation}
where
\begin{equation}
F(r)=1-{\rm exp}(-r^2/2\sigma^2).
\end{equation}

\begin{figure*}
\includegraphics[scale=0.33]{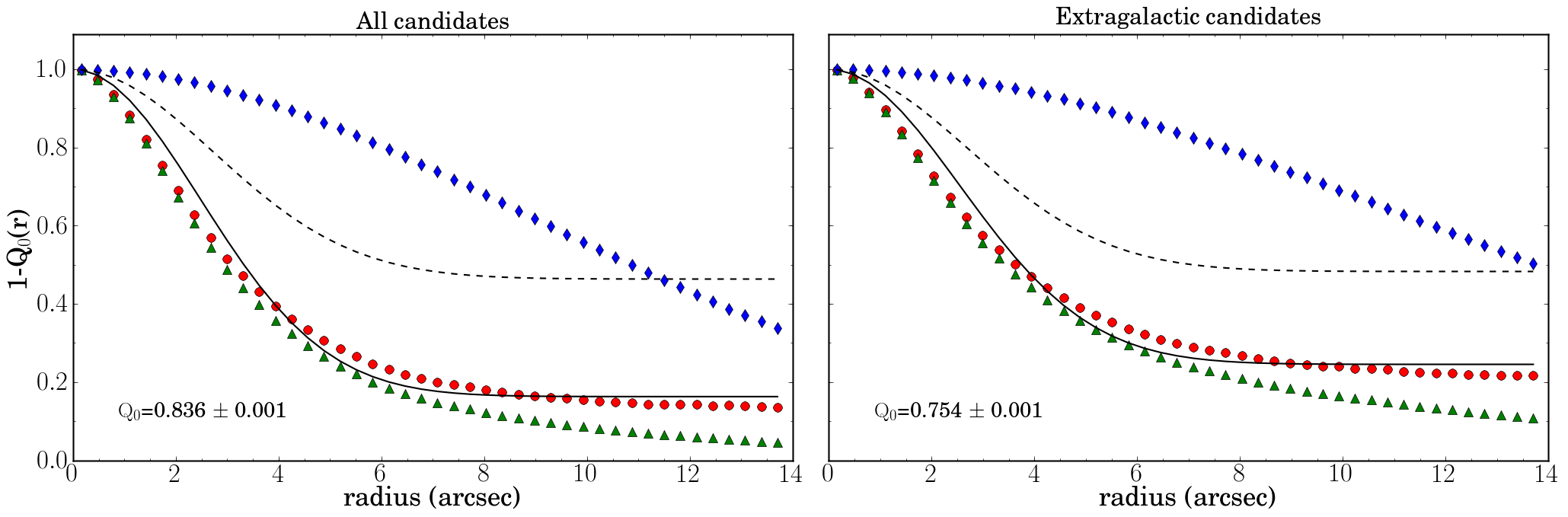}
\caption{Illustration of the method to estimate $1- Q_{\rm 0}$, the fraction of SPIRE sources without a counterpart, by counting the objects without a candidate counterpart within the search radius (blanks) as a function of the search radius. The red circles represent the values obtained by dividing the number of observed blank SPIRE positions (green triangles) by the number of blank random positions (blue diamonds). The black solid line represents the best fit to the model given in equation (\ref{q0_model}), with $Q_0$ value indicated inside the panels. Left panel shows the results for all optical candidates and right panel shows the results for extragalactic candidates only.}
\label{q0_estimate}
\end{figure*}

The number of blanks as a function of the search radius as well as the best-fitting model are shown in Fig. \ref{q0_estimate}. From the fitting we obtained $Q_0=0.519\pm0.001$ for galaxies and quasars and $Q_0=0.019\pm0.001$ for stars. This is in agreement with the results obtained in \citetalias{bourne16} for the SDSS counterparts in the GAMA fields ($Q_0=0.519\pm0.001$ for extragalactic objects and $Q_0=0.02\pm0.002$ for stars).

Note that the value of $Q_0$ estimated above is different from the one obtained previously in Sec. \ref{f_r_sec} from the fitting of the SPIRE-SDSS positional offset histogram. In the first analysis, we found a higher $Q_0$ value for the blue ($S_{250}/S_{350}> 2.4$) SPIRE sources, increasing with SNR (see Table \ref{best-fit-sigmapos}). The average $Q_0$ for the blue SPIRE sources, obtained from the number-weighted average of the values in Table \ref{best-fit-sigmapos}, is $0.783$. This difference is not only due to the fact that $Q_0$ can be biased by the clustering and by the occurrence of multiple counterparts, but mainly because $Q_0$ is a function of the colour. \citetalias{bourne16} compared different methods to estimate $Q_0$ and found that $Q_0$ can be boosted by 0.03 due to multiplicity and by 0.07 due to clustering. As in Sec. \ref{f_r_sec}, we examined the colour dependence of $Q_0$ in the same colour bins of Fig. \ref{sigma_pos_cbins}. We found a significant dependence of $Q_0$ with colour, with a variation from $Q_0=0.31$ in the reddest bin (considering only SPIRE sources with $S_{\rm 250}/S_{\rm 350}< 1.1$) to $Q_0=0.83$ in the bluest bin (SPIRE sources with $S_{\rm 250}/S_{\rm 350}>2.4$). However, the estimate of $Q_0$ for the redder sources is not reliable, as we did not account for the bias from lensing reported in \citet{bourne14}. For this reason, performing the likelihood ratio analysis as a function of colour is not straightforward and we decided to use the overall estimate of $Q_0$ to obtain our ID catalogue. 

In our likelihood ratio analysis, we also assumed a constant value of $q(m)/n(m)$ for stars with $r$-band magnitude $m<21.5$ and for extragalactic objects with $m<13.5$, following the prescription of \citetalias{bourne16} and \citetalias{Smith2011} for galaxies. This is because the $q(m)/n(m)$ distribution is not well sampled for the brightest objects. The constant value adopted corresponds to the average of $q(m)/n(m)$ within this magnitude range, as shown by the dashed line in Fig. \ref{mag_dist}. As discussed in \citetalias{bourne16}, one of the implications of this choice, in contrast with the assumption of a constant $q(m)$ for stars as in \citetalias{Smith2011}, is that the bright stars will have a likelihood ratio which is independent of the magnitude, therefore discarding the assumption of any correspondence between their optical magnitude and detection in $250\mu$m. It is also important to note that the relatively high $q(m)$ measured for stars with $m>21.5$ (see Fig. \ref{mag_dist}) can be biased by faint unresolved galaxies which were misclassified as stars. 

\subsection{Likelihood ratios and reliabilities}
\label{lr_rel_optical}

We applied the likelihood ratio method described above and obtained the likelihood ratios and reliabilities of every potential counterpart within 10 arcsec of each SPIRE source with ${\rm SNR}_{250}\ge4$. The SPIRE catalogue contains 112155 sources with ${\rm SNR}_{250}\ge4$. Of those, 77521 sources (69.1 per cent of the sample) have at least one possible optical counterpart with $r_{\rm model}<22.4$ in the SDSS selected catalogue within the search radius. In total, there are 111945 possible matches, of which 14630 (13.1 per cent) are classified as stars, 96876 (86.5 per cent) are classified as galaxies and 439 (0.4 per cent) are classified as quasars, according to the classification described in Sec. \ref{sg_sdss}. The remaining 34634 SPIRE sources (30.9 per cent of the sample) have no counterpart identified in the optical catalogue within the search radius of $10$ arcsec. Most of these sources have red SPIRE colours and possibly lie at high redshift, being too faint to be detected in SDSS.  

We visually inspected the brightest 1300 SPIRE sources in our ID catalogue to ensure that no large nearby galaxies have very low reliability. One reason for this is because large galaxies are resolved in the SPIRE maps, so that their positional errors do not follow equation (\ref{f_r}), which assumes point-like sources. Another reason is because they can be deblended into several components by the automated source extraction in both optical and submm images. In such cases, the likelihood ratios and reliabilities of the optical counterparts are under-estimated. The reliability can also be under-estimated when a SPIRE source is a merger or a blend of two objects. During the visual inspection, we also found some objects that were misclassified, leading to under or over-estimated reliabilities. We therefore flagged 56 counterparts with reliability under-estimated or over-estimated and modified their reliabilities. We also flagged 20 SPIRE sources whose optical counterparts were missed due to the incompleteness of the SDSS catalogue, in particular in the masked areas close to bright stars. 

\begin{figure*}
  \centering
    \includegraphics[scale=0.32]{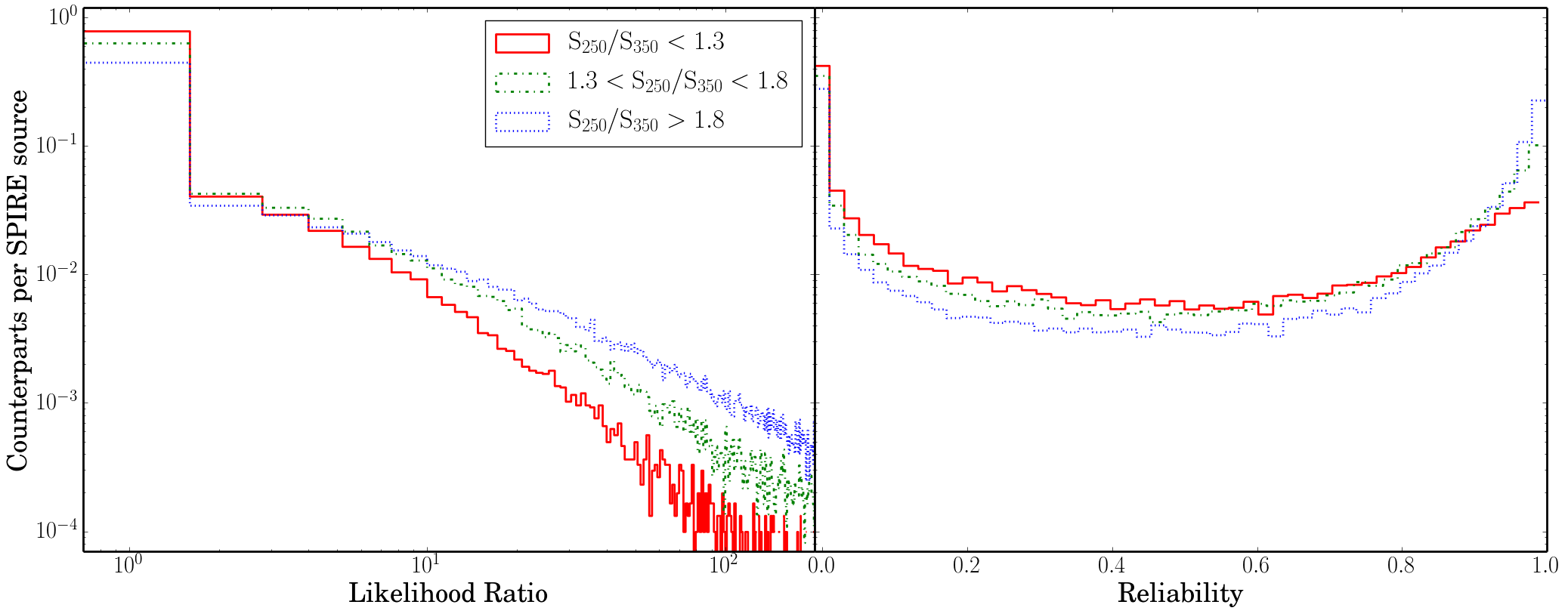}
  \caption{Likelihood ratios and reliabilities of the optical counterparts in three bins of SPIRE sources colour. }
\label{lr_rel}
\end{figure*}

The likelihood ratios and reliabilities of all potential matches are shown in Fig. \ref{lr_rel} in three bins of $S_{250}/S_{350}$-$\mu$m colour. The plots show that the fraction of SPIRE sources with higher likelihood ratios and reliabilities is larger for the blue ($S_{250}/S_{350}>1.8$) SPIRE sources than for the red ($S_{250}/S_{350}<1.0$) ones. This can be explained by the fact that red sources are more likely to be at higher redshifts and therefore less likely to be detected in SDSS. 

We consider those sources with $R_j\ge 0.8$ as being reliable counterparts. We identified 42429 reliable counterparts, which means that we were able to match 37.8 per cent of the SPIRE sources with a high reliability. Of those, 403 (1 per cent) are classified as stars, 41733 (98.3 per cent) as galaxies and 293 (0.7 per cent) as quasars. The mean reliability of the reliable matches is 0.953. The contamination of the reliable sample can be estimated as 
\begin{equation}
 N_{\rm false}=\sum_{i, R_i \ge 0.8} (1-R_i),
\label{nfalse}
\end{equation}
which is based on the assumption that the probability of a counterpart being false is $(1-R_i)$. Statistically, we expect there to be 2002 falsely identified objects with $R_j\ge0.8$, corresponding to 4.7 per cent of the reliable matches. The same false ID rate was obtained by \citetalias{bourne16} for the optical IDs in the GAMA fields. 

We define the completeness of the reliable sample as 
\begin{equation}
\eta=\frac{1}{Q_0}\frac{n(R\ge0.8)}{n({\rm SNR}_{250} \ge 4)}, 
\label{completeness}
\end{equation}
where $\eta=1$ means that the fraction of the reliable counterparts reaches $Q_0$, and the cleanness of the reliable sample as
\begin{equation}
C=1-\frac{N_{\rm false}}{N_{\rm SPIRE}}.
\label{cleanness}
\end{equation}

\begin{figure}
  \centering
    \includegraphics[scale=0.35]{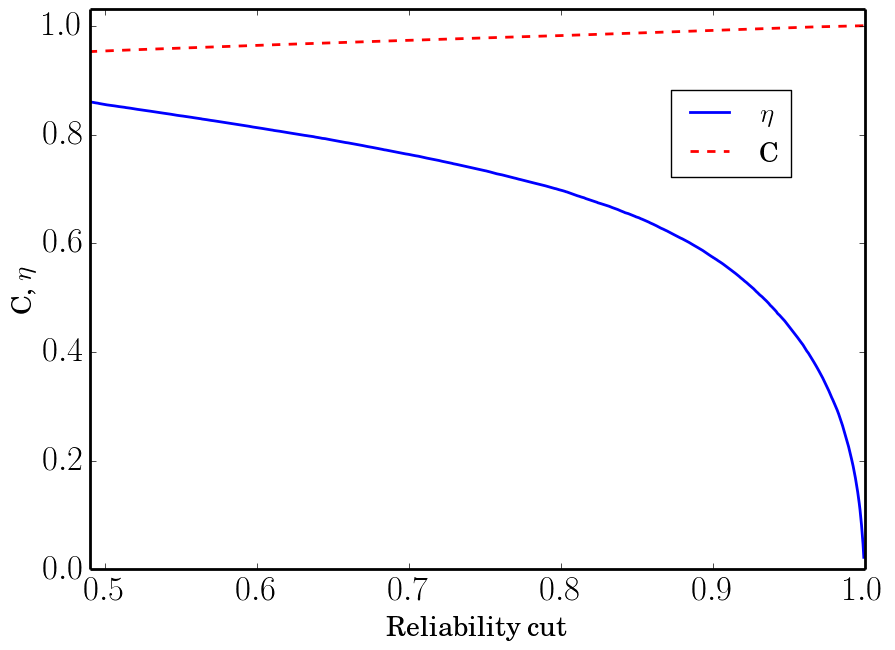}
  \caption{Completeness $\eta$ and cleanness C of the optical counterpart sample as a function of the reliability cut. }
  \label{completeness_cleanness}
\end{figure}

For the reliable extragalactic counterparts we obtained $\eta=71.7$ per cent, which is similar to the completeness of the IDs in the GAMA fields (73 per cent) obtained by \citetalias{bourne16}. In Fig. \ref{completeness_cleanness} we show how completeness and cleanness change as a function of the reliability cut. The figure shows that the reliability cut chosen in equations (\ref{completeness}) and (\ref{cleanness}), $R\ge0.8$, corresponds to a good compromise between cleanness and completeness. A higher reliability cut would imply a modest increase in the cleanness at the cost of a drastic reduction of the completeness, while a lower cut would not result in a significant increase of the completeness. 

\subsection{Multiplicity of the counterparts}
\label{multiplicity_optical}

The likelihood method assigns reliabilities in a self-consistent way, in order to deal with the case of multiple counterparts. The method assumes that there is only one true counterpart and that the sum of reliabilities of multiple candidates cannot exceed unity. This means that there is a bias against multiple genuine counterparts, such as merging galaxies or members of the same cluster, resulting from the fact that multiple counterparts in the search radius reduce the reliability of the true counterpart. For example, in the case of multiple candidates, if one of them has $R_j>0.2$, it is not possible to find a reliable counterpart. 

In Table \ref{multiplicities} we present the distribution of the number of SPIRE sources as a function of the multiplicity (m$_{\rm id}$) of the counterparts with $r_{\rm model}<22.4$ within the 10-arcsec search radius. We also show the fraction of those with reliable counterparts. For instance, we found that  50168 SPIRE sources have only one counterpart in the optical catalogue, of which 54 per cent are reliable, and that 21370 have two potential IDs, of which 57 per cent are reliable. The fraction of reliable counterparts falls slowly with increasing multiplicity, except for the last two bins. However, the number of sources in the last two bins are so small that they are statistically insignificant. The results of this table reflect the incompleteness of the ID sample due to multiplicity.

\begin{table}
\begin{center}
\begin{tabular}{ c l l l    }
\hline
m$_{\rm id}$   &      N (SPIRE)  &  \multicolumn{2}{ c }{N (reliable)} \\
\hline
0  &        34634    &  0       &(0\%)     \\
1  &        50168    &  27082   &(54\%)    \\                                                                                                           
2  &        21370    &  12155   &(57\%)   \\                                                                                                            
3  &        5032     &  2740    &(54\%)    \\
4  &        837      &  396     &(47\%)    \\          
5  &        97       &  47      &(48\%)    \\
6  &        12       &  5       &(42\%)    \\
7  &        4        &  3       &(75\%)    \\
8  &        1        &  1       &(100\%)   \\
\hline
\end{tabular}
\end{center}
\caption{Number of SPIRE sources as a function of the multiplicity (m$_{\rm id}$) of candidate IDs within the 10-arcsec search radius, counting either all candidates and reliable ones.}
\label{multiplicities}
\end{table}

Redshift information could be used to confirm whether multiple counterparts are physically associated, either as interacting systems or members of the same cluster. However, such investigation is difficult, due to the large errors in the photometric redshifts and because only 1.4 per cent of the objects in our optical catalogue have spectroscopic redshifts measured.

Another approach to avoid the effects of multiplicity is to use the likelihood ratio values instead of the reliabilities, as described in \citetalias{Smith2011} and \citetalias{bourne16}. From equation (\ref{reliability}), it is easy to demonstrate that for the cases where there is only one extragalactic counterpart the threshold $R_j\ge 0.8$ corresponds to a cut in likelihood ratio of 1.924. We found 46220 possible extragalactic counterparts with likelihood ratios above that threshold, of which 41813 also satisfy the $R_j\ge 0.8$ threshold. The remaining 4407 (9.5 per cent) counterparts that fail these reliability criteria could be considered as missed candidates of SPIRE sources with multiple counterparts. This fraction is smaller than the one obtained by \citetalias{bourne16} (13.0 per cent) using the same approach. However, those missed counterparts could be either genuine multiple associations or chance alignments. 

Alternatively, we can estimate the number of missed multiple counterparts by following the prescription from \citetalias{Fleuren2012}, which assumes that candidate matches are all associated to the same SPIRE source if the sum of their reliabilities exceeds the threshold $R=0.8$. We found that 2449 SPIRE sources have multiple extragalactic counterparts whose sum of their reliabilities exceeds the threshold $R=0.8$, but with no individual counterpart meeting the threshold. That means that those potentially true multiple counterparts are missed by applying a threshold in the reliability of individual matches. 

\subsection{Redshift distribution}

\begin{figure}
  \centering
    \includegraphics[scale=0.35]{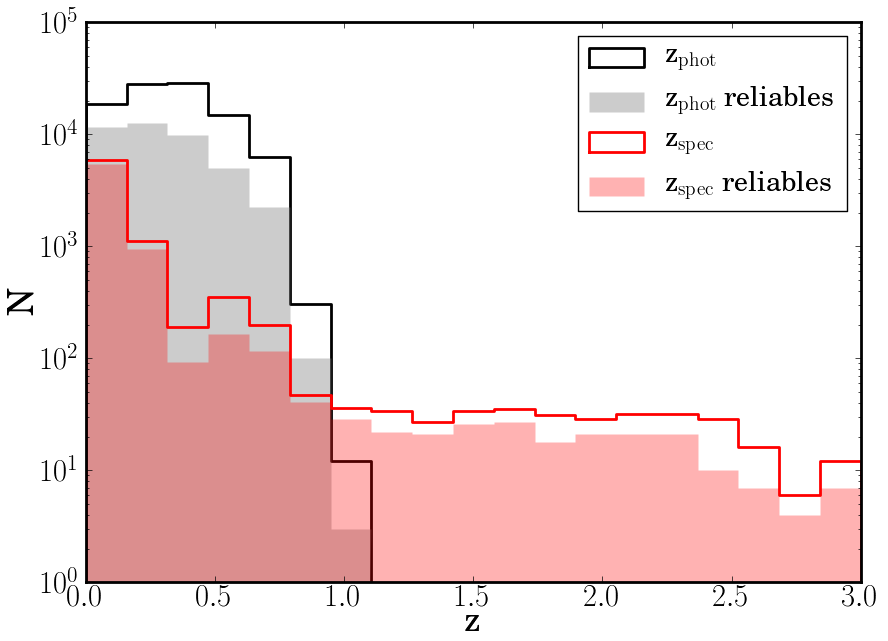}
  \caption{Redshift distribution of the optical counterparts to the SPIRE sources. The solid lines represent the redshift distribution for all potential counterparts, while the filled histograms represent the distribution for the reliable counterparts only. The black histogram corresponds to the photometric redshift distribution and the red histogram to the spectroscopic redshift distribution. }
  \label{redshift_dist_optical}
\end{figure}

The redshift distribution of all potential counterparts and reliable counterparts is show in Fig. \ref{redshift_dist_optical}. Of the 111945 potential optical counterparts, 8175 (7.3 per cent) have spectroscopic redshifts. This fraction increases by a factor of 2 for the reliable counterparts: 6975 of the 42429 (16.4 per cent) reliable counterparts have a spectroscopic redshift, all of which originate from the SDSS DR10. Only 30 reliable counterparts have a spectroscopic redshift with quality $Q_z < 3$. The larger fraction of objects in the ID catalogue with spectroscopic redshift if compared to the input optical catalogue can be explained by the fact that the LR method is more likely to find a reliable that is brighter and the brighter objects are more likely to be targeted by SDSS to have their spectroscopic redshift measured. 

From the comparison of photometric and spectroscopic redshifts distributions in Fig. \ref{redshift_dist_optical}, it can be seen that the majority of our potential counterparts have $z_{\rm phot} < 1$, as expected from the magnitude limit of our SDSS catalogue. The figure also shows that most of the IDs with a spectroscopic redshift that is higher than 1 are reliable. 

A more detailed discussion of the redshift distribution of the optical counterparts to the H-ATLAS sources can be found in \citetalias{bourne16}, since the GAMA fields have more spectroscopic redshift data.

\section{Near-infrared counterparts to submm sources}
\label{sec_id_ir}

Near-infrared wavelengths are better suited to study galaxies at high redshift than optical wavelengths, since the rest-frame UV and visible bands are shifted to the infrared part of the spectrum. The near-infrared is also less attenuated by dust which facilitates detection of the dusty and star-forming sources found by {\it Herschel}. 
We therefore expect to identify a higher number of reliable counterparts to the SPIRE sources by matching with a near-infrared catalogue than with SDSS for a given source number density. 

For this reason, we have also investigated the identification of near-infrared counterparts to the SPIRE sources with $250~\mu$m SNR$\ge4$ in a smaller area within the NGP field, using deeper data obtained with UKIRT. Although the entire NGP field is covered by the shallower UKIDSS-LAS survey and therefore the optical identifications from the previous section have NIR photometry obtained from this survey, only in this small area observed by UKIRT the data are deep enough for an independent ID analysis. Comparing the counterparts found in this deeper dataset with those in the shallower optical and near-infrared data allows additional quantification of the efficiency of the identification process as well as an understanding of the properties of the sources not identified in the fainter data.

\subsection{Near-infrared data}
\label{near-ir-data}

\begin{figure}
  \centering
    \includegraphics[scale=0.35]{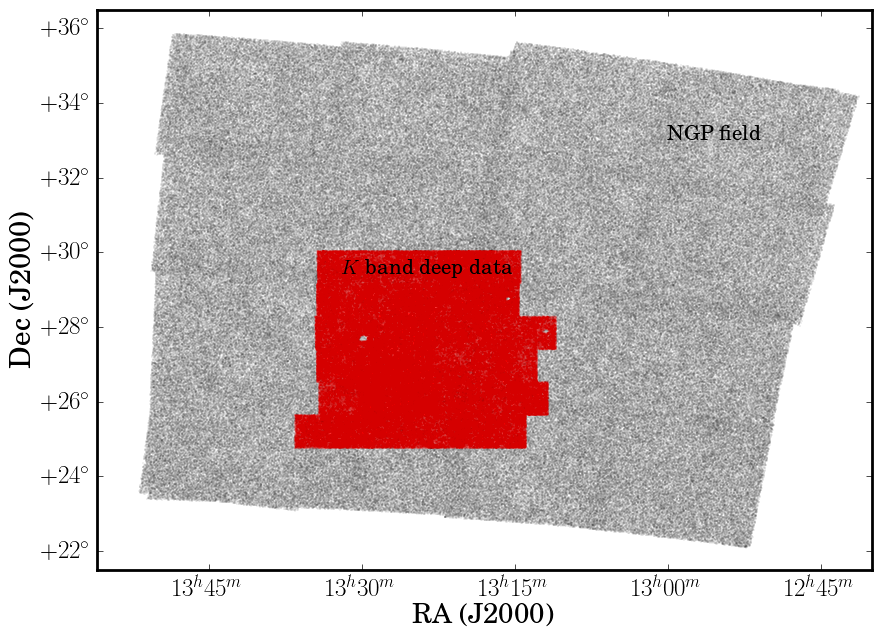}
  \caption{The $25.93$\,deg$^2$  area covered by the UKIRT/WFCAM observations in $K$ band is shown in red. The NGP field area is shown in grey.}
  \label{field_ukirt}
\end{figure}

We obtained deeper near-infrared imaging in a smaller field of size $25.93$\,deg$^2$ using the Wide Field Camera (WFCAM) on the United Kingdom InfraRed Telescope \citep[UKIRT;][]{casali07}. The field is centered approximately on the coordinates RA=13h21m and Dec=27d21m and is displayed in Fig. \ref{field_ukirt}. The observations comprise 136 pointings, resulting in 34 tiles. The total exposure time of each tile was 640 s, made up of 64 exposures of 10 s. The seeing of the individual exposures ranges from $0.43$ to $1.21$ arcsec, with an average of $0.72$ arcsec.

The raw images were processed with the WFCAM pipeline in exactly the same way as for the UKIDSS LAS \citep{Dye03}. This stacks four sets of 10 s observations belonging to one dither pattern into stacks of 40 s, then four of these are stacked into a stack of 160 s. We visually inspected the 160 s stacks and discarded those with poor data quality before stacking them into 640 s or 320 s tiles. As a result, 3 tiles (12 of the 136 pointings) have a total exposure time of 320 s.

The stacked tiles were then mosaiced together into a single image using {\sc SWarp} \citep{bertin02}. When coadding the tiles to create the mosaic, we used the normalized inverse variance weight maps provided by the WFCAM pipeline and we scaled the fluxes to take into account that some tiles have different exposure times. 

Source detection and photometry were performed using {\sc SExtractor} \citep{sextractor}. Sources with more than 6 contiguous pixels  whose flux is $1.4 \sigma$ above the sky noise were detected on the coadded image. By creating our own catalogue instead of concatenating the catalogues produced by the WFCAM pipeline for the individual stacked images we avoided duplicated objects in the overlapping areas and false detections on the edges of stacks due to noise. In addition, we were able to obtain size measurements, that are useful for the star-galaxy separation and likelihood ratio analysis. 

The near-infrared deep object catalogue contains 806630 objects. We visually inspected the mosaic image in order to remove spurious objects from the catalogue. We inspected objects with {\sc SExtractor} deblend flags and removed 150 objects corresponding to erroneous deblending. We also masked saturated stars with significant diffraction patterns and a variety of image defects that give rise to spurious detections in the catalogue. We removed 4121 objects within the masked regions. Therefore, the cleaned catalogue that was used as input object catalogue in the likelihood ratio analysis contains 802359 objects. Since the limiting magnitude can slightly vary across the field, we adopted the shallowest limiting magnitude ($5\,\sigma$ detection limit) of $K=19.4$ (Vega system).

We assessed the quality of our $K$-band photometry using the well-calibrated data from UKIDSS-LAS. We performed an object matching (3 arcsec search radius) and compared our 2-arcsec aperture magnitudes with those from UKIDSS-LAS. Considering only the UKIDSS-LAS objects with $K_{\rm LAS}<18.4$ and after applying a $2.5\sigma$ clipping to eliminate outliers, we obtained the mean offset $\langle K-K_{LAS}\rangle=0.003$ with an {\it rms} value of $0.09$. Approximately $8$ per cent of the cross-matched sources have large differences in their $K$-band magnitudes and were clipped as outliers. The majority of them are in the faint end of the magnitude distribution, where the magnitude errors are larger. Moreover, the mean positional offsets of the outliers is $17$ per cent larger than for the objects which satisfied the $2.5\sigma$ clipping, indicating that a fraction of the outliers could be the wrong match within the search radius. These results suggests that our photometric calibration is reliable. 

The near-infrared deep object catalogue is one of the products of H-ATLAS DR2.

\subsubsection{Star-galaxy separation}

\begin{figure}
  \centering
    \includegraphics[scale=0.35]{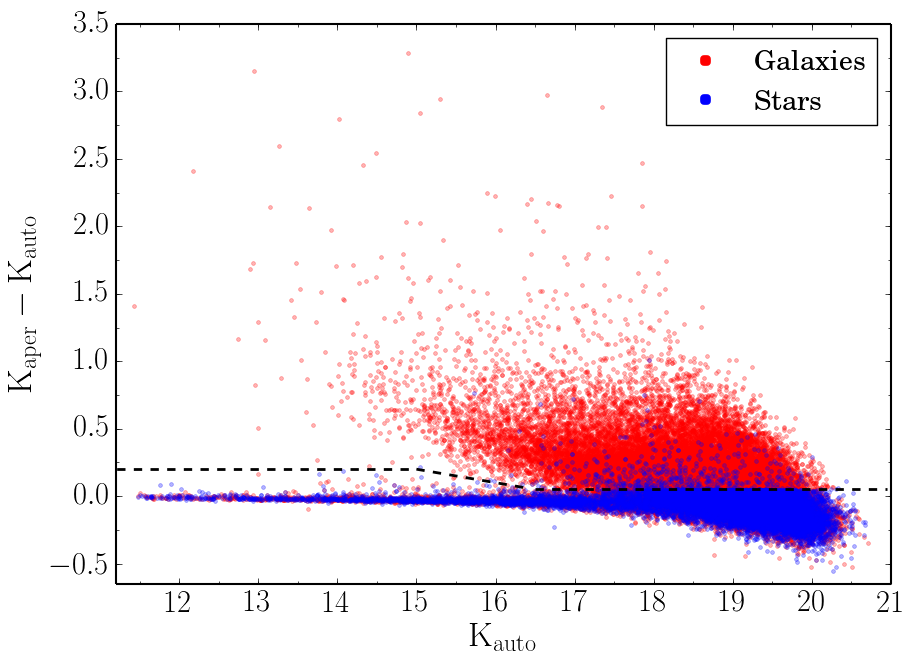}
  \caption{Star-galaxy separation for the near-infrared input catalogue. For objects with a match in SDSS and UKIDSS-LAS, the classification described in Sec. \ref{optical_data} was used. For the remaining objects, we use the criteria described by equations (\ref{sgsep_k}) and (\ref{fsg}) shown as a dashed line, which is based on the fraction of extended flux in $K$ band and size measurements.}
\label{sgsep_ukirt}
\end{figure}

Given the lack of colours in our NIR deep object catalogue, we were not able to follow a similar prescription as described in Sec. \ref{sg_sdss} to separate star and galaxy populations. However, we were able to use the results of star-galaxy separation obtained previously for the optical catalogue (based on color and shape parameters) for the objects in our NIR deep object catalogue that have a match in SDSS and UKIDSS-LAS. We matched our NIR deep object catalogue with the optical object catalogue used in Sec. \ref{optical_data} and assigned the star-galaxy flag from the latter to the matched near-infrared objects. For the remaining objects in the $K$-band catalogue, the star-galaxy separation was performed based only on the comparison of the morphology of objects with the morphology of point-like objects, as illustrated in Fig. \ref{sgsep_ukirt}. In this figure we plot $\Delta_{\rm K}$ as a function of $K_{\rm auto}$, where $\Delta_{\rm K}$ is defined as 
\begin{equation}
 \Delta_{\rm K} = K_{\rm aper} - K_{\rm auto},
 \label{delta_k}
\end{equation}
$K_{\rm aper}$ is the 2-arcsec aperture magnitude and $K_{\rm auto}$ is the automatic aperture magnitude. 

Fig. \ref{sgsep_ukirt} shows that for bright magnitudes the extended flux of galaxies makes $\Delta_{\rm K}$ strongly bimodal and the stellar locus is easily separated. For fainter magnitudes, as galaxies become unresolved, the galaxy peak moves down and merges with the stellar peak. Note the absence of saturated stars in the bright end of Fig. \ref{sgsep_ukirt}, as those objects were masked and removed from the input near-infrared catalogue. 

Based on the results of Fig. \ref{sgsep_ukirt}, for the objects in our NIR deep object catalogue without a SDSS match, galaxies are defined as objects satisfying the following criteria
\begin{align}
&\Delta_{\rm K} > 0.2 {\rm \,\,and\,\,} \mbox{FWHM} >0.8'' \nonumber \\
&\mbox{or}\nonumber \\
&\Delta_{\rm K} > f_{sg}(K_{\rm auto}) \mbox{~~and~~FWHM} >0.8'',
\label{sgsep_k}
\end{align}
where FWHM is the full-width at half maximum computed by {\sc SExtractor} (assuming a Gaussian profile for the object), and 
\begin{equation}
f_{\rm sg}(x)  = 
\begin{cases}
0.2 -\frac{1}{10}(x-15), &  15.0< x <16.5\\
0.05,  &  x>16.5.\\
\end{cases}
\label{fsg}
\end{equation}

The above criteria reproduce approximately $89$ per cent of the star-galaxy classifications obtained previously from the optical object catalogue for objects with a match in SDSS and UKIDSS-LAS. For consistency, we applied the FWHM $>0.8$ arcsec criteria for galaxies to the objects whose star-galaxy flag was obtained from the optical object catalogue to update their classification.

In Fig. \ref{sgsep_ukirt} we show the separation criteria used from equations (\ref{sgsep_k}) and (\ref{fsg}). Approximately $45$ per cent of the objects were classified as stars. Note that part of the contamination of the galaxy and stellar samples in this figure comes from the misclassifications in the optical object catalogue. 

\subsection{Likelihood ratio analysis}
\label{LR_ukirt}

We applied the LR technique described in Sec. \ref{SecLR} to identify the most reliable near-infrared counterparts to the 250-$\mu$m SPIRE sources. In the following we summarize the results, pointing out any modification to the previous approach. 

\begin{figure*}
  \centering
    \includegraphics[scale=0.35]{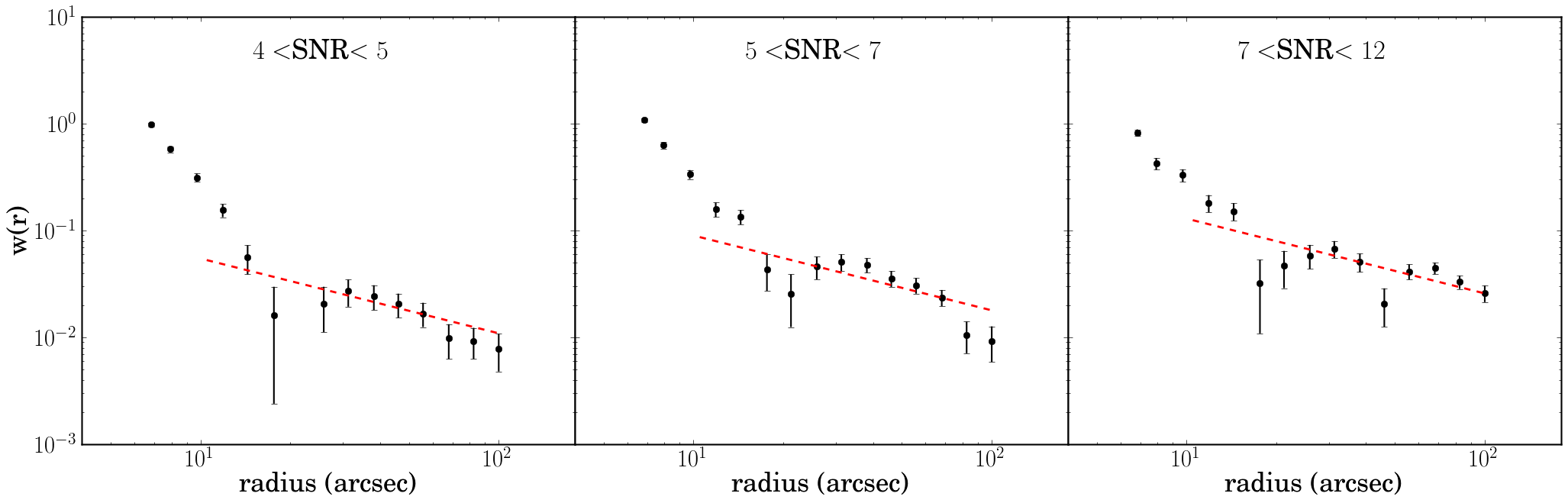}
  \caption{Cross-correlation between the SPIRE $250~\mu$m positions and the UKIRT $K$ band position is bins of SPIRE SNR. Error bars represent Poisson errors. The fit was performed with power-law slope fixed to $-0.7$ and only for $r>10$ arcsec. The best-fitting correlation length ($r_0$) are given in Table \ref{best-fit-sigmapos-ukirt}.}
\label{crosscorr_ukirt}
\end{figure*}

We begin by estimating the width $\sigma_{\rm pos}$ of the positional error distribution $f(r)$. As in the Sec. \ref{f_r_sec}, we fitted the two-dimensional histogram of $\Delta \rm{RA}$ and $\Delta {\rm Dec}$ separations with the model described by equation \ref{histogram_distribution}, which includes the contribution of the true counterparts, the random background of chance alignments and a galaxy clustering term. Given the new data set, we no longer can use the results obtained by \citetalias{bourne16} for the cross-correlation function. We therefore measured the angular cross-correlation function between the SPIRE sources and near-infrared objects in different SNR bins. We used a modified version of the Landy-Szalay estimator \citep{landy93},
\begin{equation}
w(r)=\frac{D_1D_2(r)-D_1R(r)-D_2R(r)+RR(r)}{RR(r)},
\label{ls-estimator}
\end{equation}
which is a linear combination of ratios between pair counts of SPIRE ($D_1$) and NIR ($D_2$) data and/or random catalogues ($R$), as a function of the radial separation $r$. The results of the cross correlation are shown in Fig. \ref{crosscorr_ukirt} for three 250-$\mu$m SNR bins. We modelled $w(r)$ as a power-law given in equation (\ref{wr}), as a function of $r$ up to 100 arcsec. We fixed the power-law index as $\delta=-0.7$ and excluded the data at $r<10$ for the fitting, as at small radii the pair counts are biased due to the presence of true counterparts. The best-fitting models for each SNR bin are shown by the dashed lines in Fig. \ref{crosscorr_ukirt}. The best-fitting values for the correlation length $r_0$ do not differ significantly from the ones obtained in the optical analysis and they are presented in Table \ref{best-fit-sigmapos-ukirt}. 

\begin{figure}
  \centering
    \includegraphics[scale=0.30]{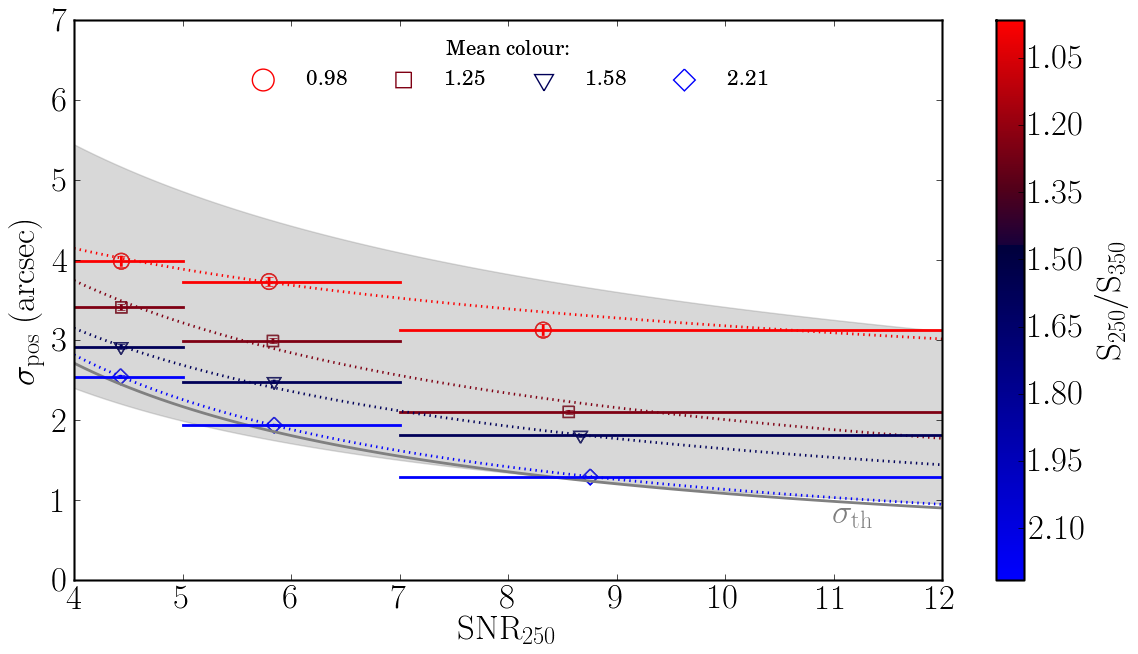}
  \caption{$\sigma_{\rm pos}$ measured from the fitting of the two-dimensional offset histogram as a function of the mean value of $250~\mu$m SNR in each SNR bin for four bins of $S_{\rm 250}/S_{\rm 350}$ colour. The symbols indicate the colour bins. The solid line corresponds to the theoretical prediction for $\sigma_{\rm pos}$ given in equation (\ref{sigma_pos_the}). Dotted lines show the best-fitting models for $\sigma_{\rm pos}$ as a function of SNR for each colour bin. The results for bluest colour bin are summarized in Table \ref{best-fit-sigmapos-ukirt}. The filled area delimits the best-fitting values of $\sigma_{\rm pos}$ measured using the optical object catalogue (Fig.\ref{sigma_pos_cbins}). The colour dependence of $\sigma_{\rm pos}$ is weaker in the NIR than in the optical (see Sec. \ref{comparison} for further discussion). }
\label{sigma_pos_cbins_ukirt}
\end{figure}

The results of the fitting of positional offsets between near-infrared data and SPIRE sources as a function of 250-$\mu$m SNR and colour are shown in Fig. \ref{sigma_pos_cbins_ukirt}. The best-fitting parameters in modelling the near-infrared positional offsets to blue ($S_{\rm 250}/S_{\rm 350}>2.4$) SPIRE sources are shown in Table \ref{best-fit-sigmapos-ukirt}. Given the smaller data set, we reduced the number of bins of SPIRE SNR and colours in comparison to the analysis of the optical counterparts. As in the optical analysis, we obtained a broader positional offset distribution for the redder sources, which is attributed to the lensing bias \citep{bourne14}. However, this effect is weaker than in the optical, as it can be seen by the shaded area in the figure, which indicates the limits of the best-fitting models for $\sigma_{\rm pos}$ obtained in the optical analysis. The positional error of redder SPIRE sources in the NIR is smaller than the one measured using the optical catalogue.

The dependence of the positional error on the $250~\mu$m SNR is modelled as the power-law function of equation (\ref{sigma_pos_snr}). The dotted lines in Fig. \ref{sigma_pos_cbins_ukirt} correspond to best-fitting models for each colour bin. For the bluest sources we obtained that $\sigma (5)=2.26 \pm 0.11$ and the slope of the power-law is $\alpha=-0.99 \pm 0.04$. This empirical dependence of $\sigma_{\rm pos}$ on SNR for the blue SPIRE sources is very close to the theoretical prediction of equation (\ref{sigma_pos_the}), which is plotted as a grey solid line figure.

\begin{table}
\begin{center}
\begin{tabular}{ c c c c c }
\hline
SNR & $N_{SPIRE}$ & $r_{0}$ (arcsec) & $\sigma_{pos}$ (arcsec) & $Q_0$ \\
\hline
4 - 5 & 1822 &  0.158 $\pm$ 0.02  & 2.547 $\pm$ 0.03 &  1.000 $\pm$ 0.000\\
5 - 7 & 1335 &  0.320 $\pm$ 0.03  & 1.941 $\pm$ 0.02 &  1.000 $\pm$ 0.000\\
7 - 12 & 735 &  0.539 $\pm$ 0.04  & 1.293 $\pm$ 0.02 &  0.917 $\pm$ 0.011\\
\hline
\end{tabular}
\end{center}
\caption{Best-fitting parameters in modelling near-infrared positional offsets to blue SPIRE sources ($S_{\rm 250}/S_{\rm 350}>2.4$) in bins of $250~\mu$m SNR.}
\label{best-fit-sigmapos-ukirt}
\end{table}

\begin{figure}
\includegraphics[scale=0.35]{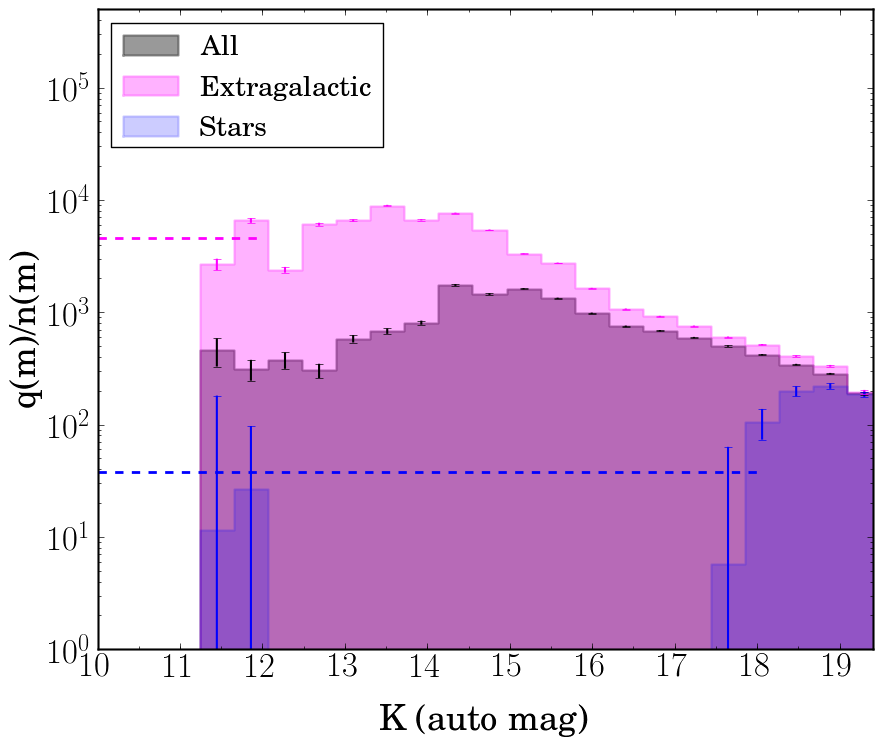}
\caption{The $K$-band magnitude distribution $q(m)/n(m)$ for extragalactic objects and stars. We assume a constant value corresponding to the average of $q(m)/n(m)$ at magnitudes $m<18$ for stars and $m<12$ for extragalactic objects, since at brighter magnitudes the distributions are not well sampled.}
\label{mag_dist_ukirt}
\end{figure}

We then measured the $K$ magnitude distributions $n(m)$ and $q(m)$ for the near-infrared counterparts, following the prescription of Sec. \ref{n_q}. The {\sc SExtractor} automatic aperture magnitude (MAG\_AUTO) was adopted in this case. In Fig. \ref{mag_dist_ukirt} we show the magnitude distribution $q(m)/n(m)$, which is used in the likelihood ratio analysis (equation \ref{LR}). As for the optical case, the value of $q(m)/n(m)$ for brighter magnitudes ($K<18$ for stars and $K<12$ for extragalactic objects) is fixed at the average within that range.

\begin{figure*}
\includegraphics[scale=0.33]{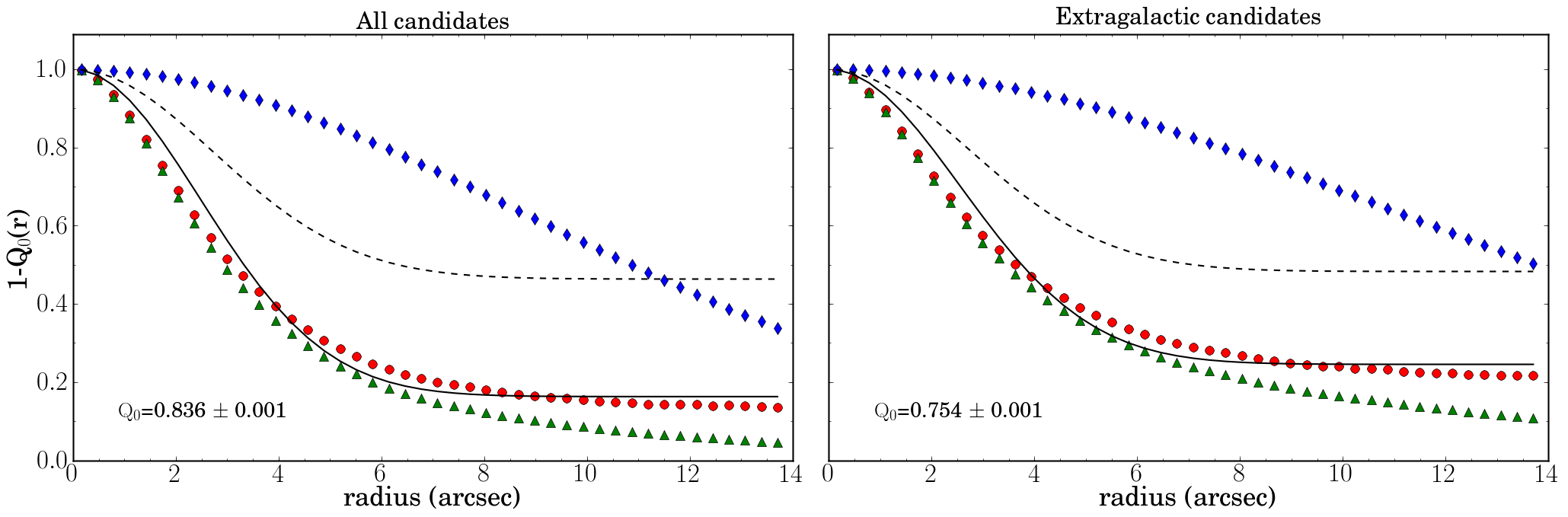}
\caption{The method of determining the fraction of SPIRE sources without a counterpart ($1- Q_{\rm 0}$), by counting the objects with no candidate counterpart with $K<19.4$ within the search radius as a function of the search radius. The red circles represent the values obtained by dividing the number of observed blank SPIRE positions (green triangles) by the number of blank random positions (blue diamonds). The black solid line represents the best fit to the model given in equation \ref{q0_model}, with $Q_0$ value indicated inside the panels. Left panel shows the results for all optical candidates and right panel shows the results for extragalactic candidates only. Dashed lines correspond to the best-fitting model obtained for the optical case.}
\label{q0_estimate_ukirt}
\end{figure*}

We estimated the normalization of the probability distribution $q(m)$, $Q_0$, by counting blanks (sources without a counterpart candidate) as a function of the search radius and fitting the model of equation (\ref{q0_model}). In Fig. \ref{q0_estimate_ukirt} we present the results of the method.  We measured $Q_0=0.754 \pm 0.001$ for extragalactic objects and $Q_0=0.082 \pm 0.001$ for stellar objects. This is similar to the value measured by \citetalias{Fleuren2012}, $Q_0=0.72 \pm 0.03$, using VISTA VIKING data down to $K_s=19.2$ (Vega system) for the Phase 1 GAMA9 field ($\approx 54$deg$^2$).  We ascribe the slight increase in our measurement of $Q_0$ to the increased depth of our near-infrared catalogue. The $Q_0$ value obtained for stars is much higher than the one obtained by \citetalias{Fleuren2012}, $Q_0=0.01$. This is mainly due to the contamination of unresolved galaxies in our stellar sample.  

Finally, we calculated the likelihood ratios and reliabilities of every potential counterpart with $K<19.4$ within 10 arcsec of each SPIRE source, using equations (\ref{LR}) and (\ref{reliability}), respectively. There are 17247 SPIRE sources with ${\rm SNR}_{250}\ge4$ in the area observed with WFCAM. We found 32041 possible counterparts with $K<19.4$ within a 10-arcsec search radius to 15780 SPIRE sources. Of those 7490 are classified as stars, 24488 as galaxies and 63 as QSOs. The remaining 1467 SPIRE sources without a counterpart identified in the near-infrared data are too faint to be detected in the WFCAM imaging or lie in masked areas around bright stars and image defects. 

We visually inspected the near-infrared image around the brightest sources in our ID catalogue down to $S_{250}=120$mJy (approximately 400 sources) and flagged 16 sources with reliability under-estimated or over-estimated, following the same procedure as for the optical ID catalogue (see Section \ref{lr_rel_optical}).

We identified 10668 reliable near-infrared counterparts. This means that we were able to match 61.8 per cent of the SPIRE sources with $R\ge0.8$. In comparison, \citetalias{Fleuren2012} found a reliable counterpart in the VISTA VIKING data for $51$ per cent of the SPIRE sources in the Phase 1 GAMA9 field. To test whether the increase in the fraction of reliable near-infrared counterparts is due to the increase in $Q_0$ and in the depth of the near-infrared catalogue, we recalculated the likelihood ratio using the same $Q_0$ values and magnitude limit as \citetalias{Fleuren2012}. We obtained that 55.3 per cent of the SPIRE sources have a reliable identification. The effect of increasing $Q_0$ and the depth of the catalogue is significant, but is not the only factor responsible for the improvement in the reliable fraction. According \citetalias{valiante16}, the use of a matched filter in the source extraction of NGP sources has reduced the positional uncertainties compared with the PSF filtering, as used in \citetalias{Fleuren2012}. This also improves the likelihood ratio and reliability of the identified counterparts. 

Using equation (\ref{nfalse}), we estimated that there are 477 false near-infrared counterparts with $R\ge0.8$, corresponding to a false ID rate of $4.5$ per cent. For the reliable extragalactic counterparts, from equation (\ref{completeness}), we obtained the completeness $\eta=$0.74. The contamination rates and completeness in the near-infrared are similar to those obtained in \citetalias{Fleuren2012} and in the previous optical analysis, indicating that the application of the likelihood ratio method was consistent and that the performance of the method is independent of the wavelength or depth of the input catalogue. 

In Table \ref{multiplicities_nir} we show the number of SPIRE sources matched and the fraction of reliables as a function of the multiplicity (number of candidate IDs per position, m$_{\rm id}$). For a more detailed discussion about the multiplicity and the comparison with the optical matching results, please see Sec. \ref{comparison}.

\begin{table}
\begin{center}
\begin{tabular}{ c l l l  l l l }
\hline
 &  \multicolumn{3}{ c }{Deep $K$ matching} &  \multicolumn{3}{ c }{Shallow $K$ matching} \\
 &  \multicolumn{3}{ c }{$K <19.40$} &  \multicolumn{3}{ c }{$K <18.69$} \\
\hline
m$_{\rm id}$   &      N (SPIRE)  &  \multicolumn{2}{ c }{N (reliable)}&      N (SPIRE)  &  \multicolumn{2}{ c }{N (reliable)} \\
\hline
0  &        1467    &  0       &(0\%)   &       3227    &  0      &(0\%)     \\
1  &        5884    &  48587   &(78\%)  &       7383    &  5429   &(74\%)  \\                                                                                                           
2  &        5505    &  3752   &(68\%) &        4566    &  2946   &(64\%)   \\                                                                                                            
3  &        2921    &  1626   &(56\%) &        1580    &  852    &(54\%)   \\
4  &        1081    &  534    &(49\%) &        389     &  193    &(50\%)    \\          
5  &        300     &  136    &(45\%) &        83      &  34     &(41\%)   \\
6  &        66      &  22     &(33\%) &        18      &  5      &(28\%)    \\
7  &        20      &  9      &(45\%) &        1       &  0      &(0\%)     \\
8  &        3       &  2      &(67\%) &        0       &  0      &(0\%)  \\
\hline
\end{tabular}
\end{center}
\caption{Number of SPIRE sources as a function of the multiplicity (m$_{\rm id}$) of NIR candidate IDs within the 10-arcsec search radius, counting either all candidates and reliable ones, in both deep and shallow $K$-band matching. }
\label{multiplicities_nir}
\end{table}

\begin{figure*}
  \centering
    \includegraphics[scale=0.32]{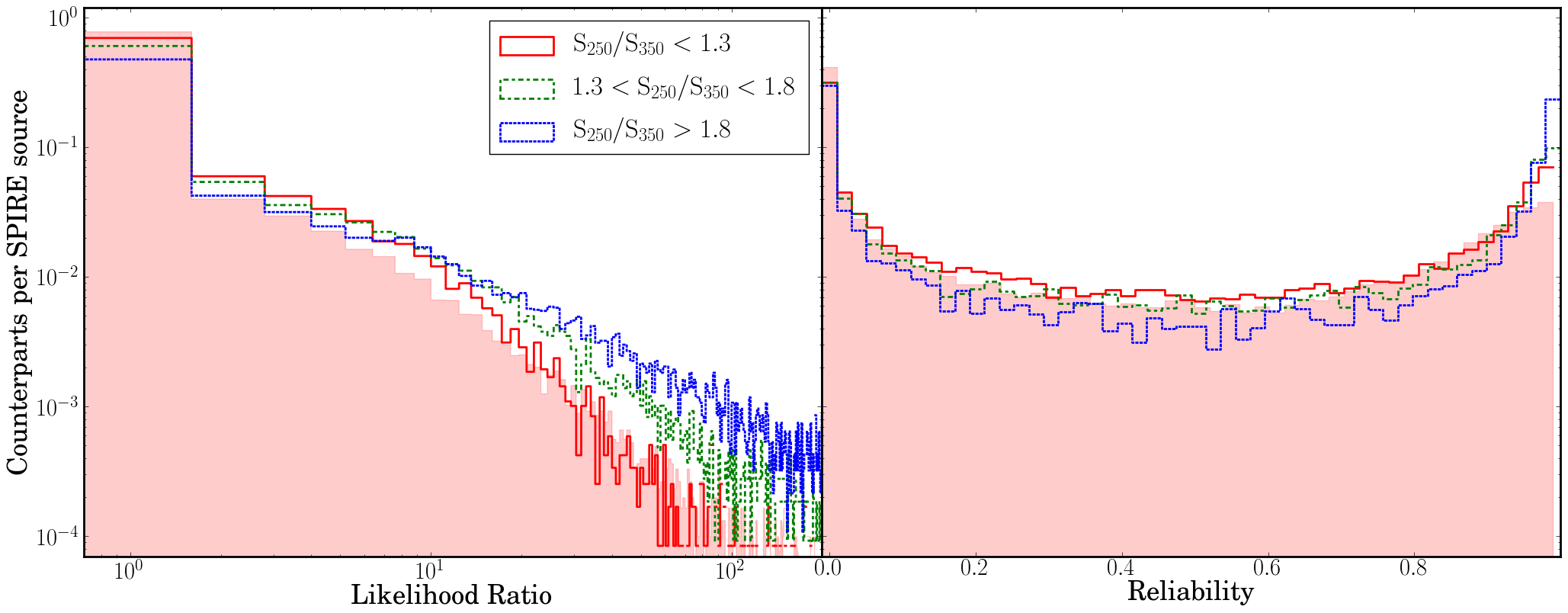}
  \caption{Likelihood ratios and reliabilities of the near-infrared counterparts with $K<19.4$ in three bins of SPIRE sources colour. The filled histograms correspond to the results of the optical counterparts to the reddest SPIRE sources.}
\label{lr_rel_ukirt}
\end{figure*}

The likelihood ratios and reliabilities are shown in Fig. \ref{lr_rel_ukirt} for three bins of SPIRE $S_{\rm 250}/S_{\rm 350}$ colour. For comparison, we also show the results of the optical counterparts to the reddest SPIRE sources (from Fig. \ref{lr_rel}) as filled histograms. We can see a significant improvement in the NIR matching of the number of highly reliable counterparts to redder SPIRE sources. 

\section{Comparing optical and NIR identifications}
\label{comparison}

In this section we investigate the performance of the likelihood ratio method to identify optical and near-infrared counterparts to the SPIRE sources. We present a comparison of the counterparts identified by the likelihood ratio method in the SDSS and in our $K$-band object catalogue. We also investigate the optical and near-infrared properties of the H-ATLAS counterparts.

In Table \ref{summary_optical_nir} we summarize the results from previous sections for the optical and NIR identifications to the SPIRE sources with SNR$_{250}\ge 4$ (second and third columns). 

The number of reliably identified sources in both optical and near-infrared matching corresponds to approximately $70$ per cent of the estimated number of true counterparts which are above the magnitude limit, given by $Q_0$. The remaining sources can be associated to one or more low-reliability match(es). Our estimates also indicate that contamination of misidentified sources ($N_{\rm false}$) is 4.7 per cent for the optical ID catalogue and $4.5$ per cent for the near-infrared ID catalogue. These estimates are lower limits, as these contamination fractions do not include the effects of lensing, for which we cannot correct in this work. Follow-up observations in radio/submm with subarcsec position are required to confirm these estimates. 

\begin{figure}
\includegraphics[scale=0.35]{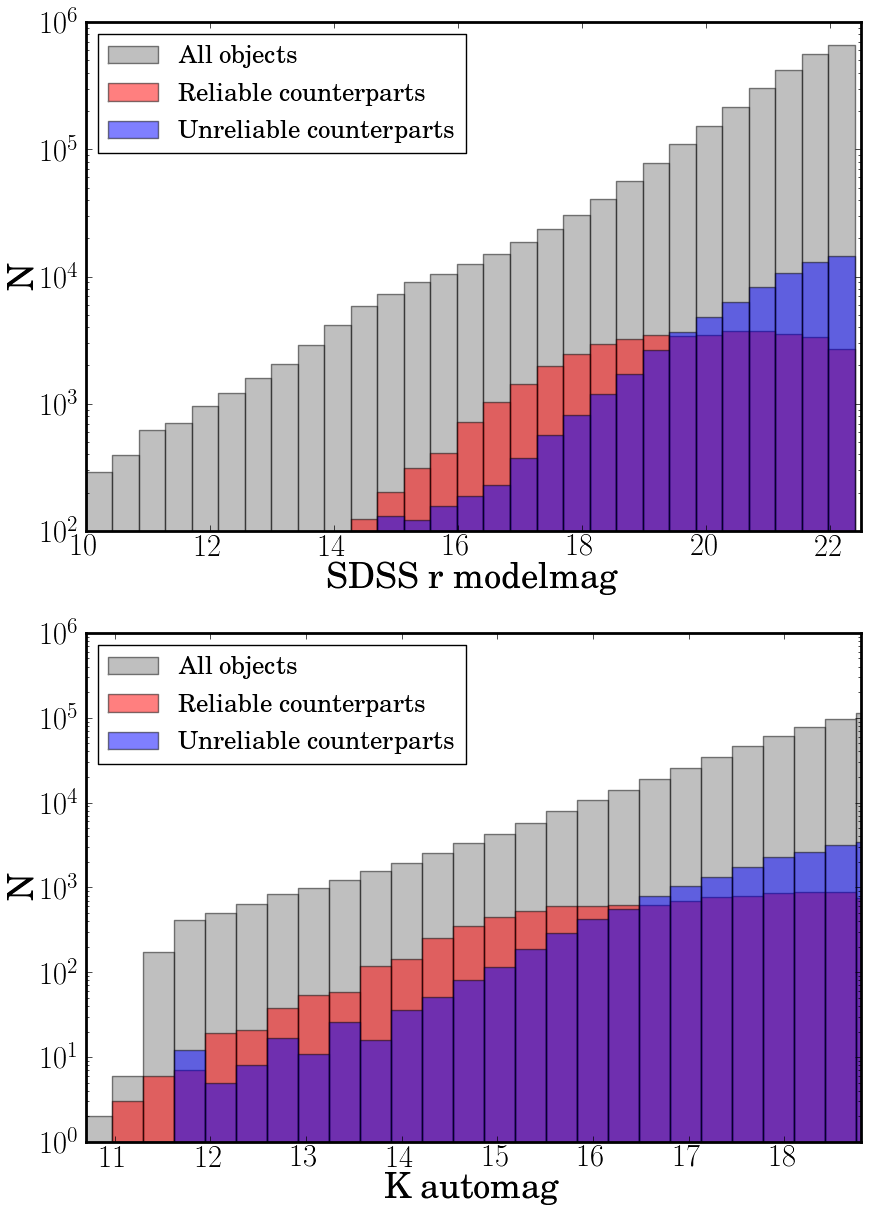}
\caption{Magnitude distribution of all objects in the input catalogue and of objects identified as reliable or unreliable counterparts of a SPIRE source in the optical (upper panel) and near-infrared (lower pannel) matching cases.}
\label{mag_dist_rel_unrel}
\end{figure}

We compare the magnitude distribution of the reliable and unreliable counterparts for the $r$-band and $K$-band matching in Fig. \ref{mag_dist_rel_unrel}. In the figure we also show the magnitude distribution of all objects in the corresponding input catalogue. We can see that in both cases the magnitude distribution of the unreliable identifications is similar to the overall magnitude distribution of the objects in the input catalogue. This indicates, as expected, that a significant fraction of unreliable counterparts may consist of unrelated background objects. 

\begin{figure}
\includegraphics[scale=0.35]{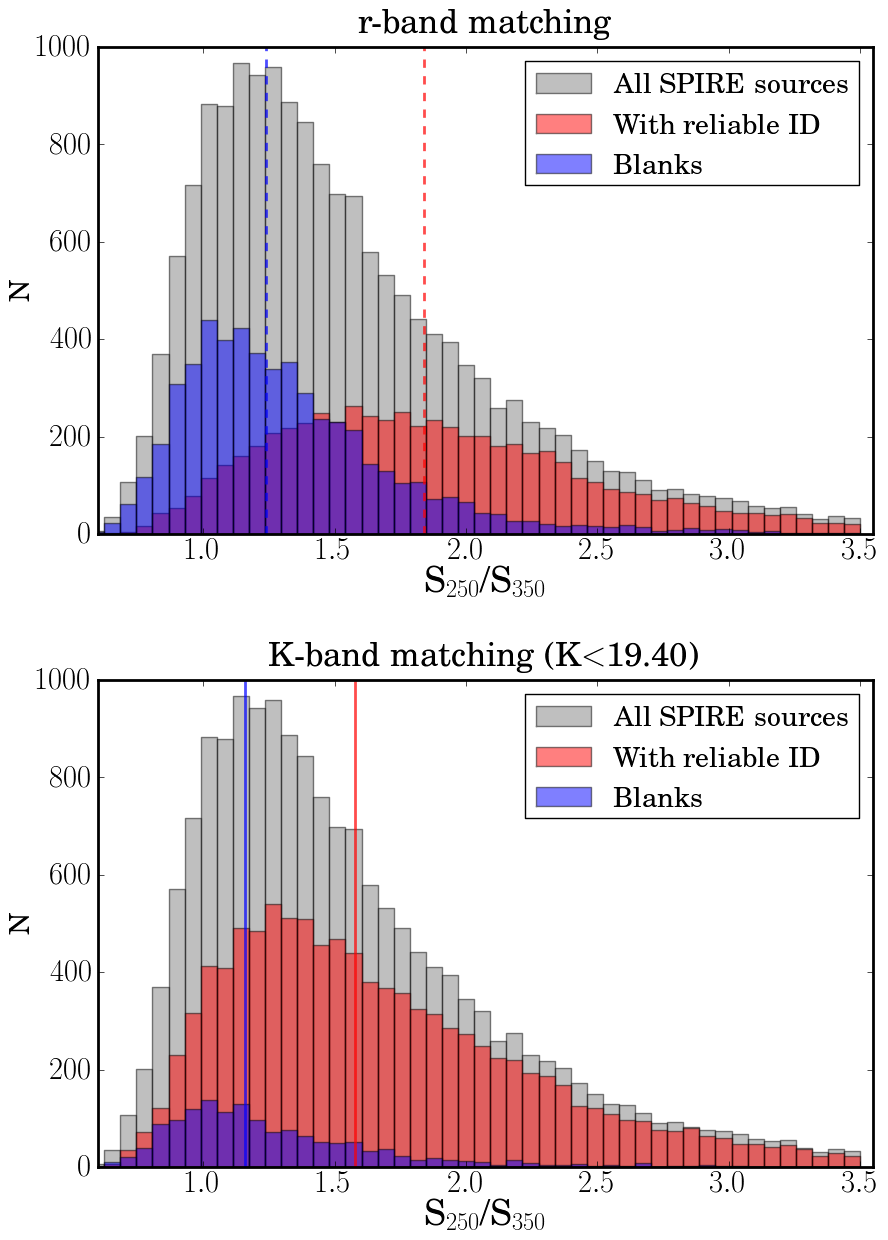}
\caption{$S_{\rm 250}/S_{\rm 350}$ colour distribution of of the SPIRE sources. The grey histogram correspond to the color distribution of all SPIRE sources with ${\rm SNR}_{250}\ge4$. The red histogram corresponds to those sources with a reliable match and the blue histogram to those with no match (blanks). Upper panel shows the results for sources in the $r$-band matching, while lower panel present the results of $K$-band matching. The vertical lines correspond to the median values of the distributions.  }
\label{colour_dist_rel_unrel}
\end{figure}

In Fig. \ref{colour_dist_rel_unrel} we compare the $S_{\rm 250}/S_{\rm 350}$ colour distribution of SPIRE sources with reliable counterparts identified to the distribution of sources with no counterpart (blanks) for both optical matching (upper panel) and near-infrared matching (lower panel). The redder colours of the blank SPIRE sources in both panels (median $S_{\rm 250}/S_{\rm 350}=1.24$ for the optical matching and median $S_{\rm 250}/S_{\rm 350}=$ 1.16 for the near-infrared matching) suggests that those are at higher redshifts and are too faint to be detected in SDSS and $K$-band catalogues. The SPIRE sources reliably identified are distinctively bluer in the optical matching (median $S_{\rm 250}/S_{\rm 350}=$ 1.84) than in the near-infrared matching (median $S_{\rm 250}/S_{\rm 350}=1.58$). In both optical and near-infrared matching, the colours of SPIRE sources with unreliable counterparts lie in between the reliable and blank populations, suggesting that this subset of sources is formed by both populations. 

The results above indicate that, as expected, our deeper $K$-band data are more effective than the SDSS r band data for reliably identifying SPIRE sources, especially the redder ones. However, a more direct comparison between the identification of counterparts in two different wavelengths should take into account the depth (and surface density of objects) of both catalogues. Moreover, the comparison should be done using similar areas, in order to minimize any differences arising from cosmic variance. Therefore we assessed the performance of the likelihood ratio to identify optical and near-infrared counterparts to the SPIRE sources considering only the optical identifications in the area overlapping to the NIR observations and using a shallower magnitude limit in $K$-band catalogue ($K<18.69$), chosen so that the object surface density matches the SDSS one. 

We applied the mask created for the WFCAM $K$-band data analysis to the optical ID catalogue obtained as described in Section \ref{lr_rel_optical} in order to have the optical identifications to the SPIRE sources within the overlapping $25.93$ deg$^2$. The number of optical counterparts, reliable counterparts and blank fields are presented in the fourth column of Table \ref{summary_optical_nir}. 
 
We then re-calculated the LR analysis for the near-infrared data using an NIR shallow object catalogue containing only objects with $K<18.69$. The results of this new analysis are summarized in the fifth column of Table \ref{summary_optical_nir}. We obtained $Q_0=0.680$ for extragalactic objects and $Q_0=0.043$ for stars. The overall value of $Q_0$ estimated for candidates in the shallower near-infrared catalogue ($Q_0=0.723$) is much higher than the one estimated for the optical counterparts in SDSS ($Q_0=0.538$), confirming the expectation that $K$ band is much better able to identify counterparts to SPIRE sources than $r$ band. The completeness of the reliable sample is $\eta=$ 0.76 and the false ID rate is 4.4 per cent. 

\begin{table*}
\begin{center}
\begin{tabular}{ l c c c c  }
\hline
 & $r$-band full area & $K$-band ($K<19.40$) & $r$-band overlap & $K$-band ($K<18.69$)\\
\hline
Area (deg$^2$)                & 177.13    &   25.93    &   25.93   &   25.93 \\
SPIRE SNR$_{250}\ge 4$        & 112155    &   17247    &   17247   &   17247\\ 
$Q_0$                         & 0.538     &   0.836    &   0.538   &   0.723 \\
Counterparts $\le 10''$       & 111945    &   32041    &   17107   &   23341 \\ 
Reliable counterparts         & 42429     &   10668    &   6285    &   9459\\ 
SPIRE with counterpart        & 77521     &   15780    &  11829    & 14020   \\ 
SPIRE blanks                  & 34634     &   1467     &  5417     & 3227 \\ 
\hline
\end{tabular}
\end{center}
\caption{Summary of results for optical and near-infrared identifications to the SPIRE sources. The second column corresponds to the results of the LR method for the full NGP area. The third column represents the results of the LR method applied to the WFCAM data, using its original estimated depth ($K<19.40$). The fourth column shows the results of the optical ID catalogue in the area overlapping to the WFCAM observations. The last column corresponds to the results of the LR method applied to the WFCAM data using a shallower limiting magnitude, chosen so that the NIR surface density matches the SDSS one ($K<18.69$).}
\label{summary_optical_nir}
\end{table*}

\begin{figure}
\includegraphics[scale=0.35]{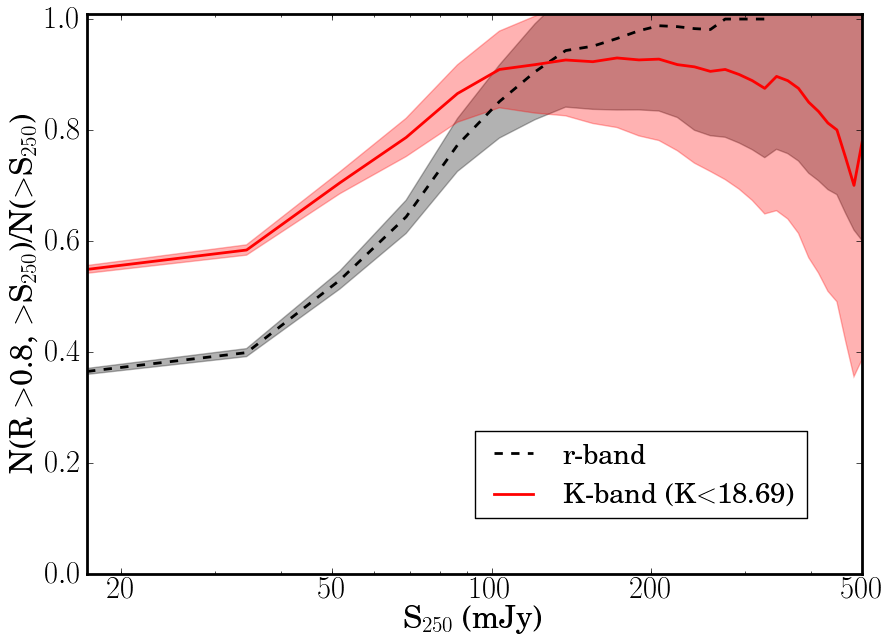}
\caption{Completeness of our optical and NIR ID catalogues (in the $25.93$ deg$^2$ overlapping area and with matched source density) as a function of $S_{\rm 250}$. The shaded regions correspond to the Poisson errors on the fractions. }
\label{completeness_flux}
\end{figure}

We computed the fractional completeness of our ID catalogues as the fraction of reliable counterparts brighter than a given flux density. In Fig. \ref{completeness_flux} we show the fractional completeness of our optical and NIR identification catalogues (in the overlapping area and with matched surface density) as a function of $250\mu$m flux density. The filled regions in the figure indicate the uncertainty on the completeness, based on the Poisson errors on the number counts. Overall, we were able to reliably identify an optical counterpart with $r<22.40$ for 36.4 per cent of our SPIRE sources in the  $25.93$ deg$^2$ overlapping area. For the case of matching to our NIR shallow object catalogue, we found a reliable counterpart with $K<18.69$ for 54.8 per cent of our $250$-$\mu$m selected sample. The figure shows that for both cases the number of missed counterparts increases for lower flux densities. This is because fainter sources are more likely at higher z and more likely to be missing from the optical and NIR surveys. 

In Table \ref{multiplicities_nir} we investigate whether there is any evidence of multiple genuine counterparts being missed in the NIR data by computing the fraction of reliables as a function of the multiplicity (number of candidate IDs per position, m$_{\rm id}$). We present the results for the multiplicity computed in both full-depth ($K <19.40$) and matched-depth ($K <18.69$) NIR matching. The variation of the reliable fraction with multiplicity is similar for the two depths. Comparing the results of this table with the values from Table \ref{multiplicities}, the fraction of reliable counterparts in the NIR decreases much faster with increasing multiplicity than in the optical case. This suggests that we are missing more multiple galaxies that are physically associated in the NIR. Since the $K$-band is much more
effective in identifying sources at high-redshift than the optical, this result might indicate that interactions of galaxies are much more important at higher redshifts. Another explanation for this result is that it is due to geometric effects, as galaxies will appear closer if they lie at higher redshifts.

\begin{figure}
\includegraphics[scale=0.35]{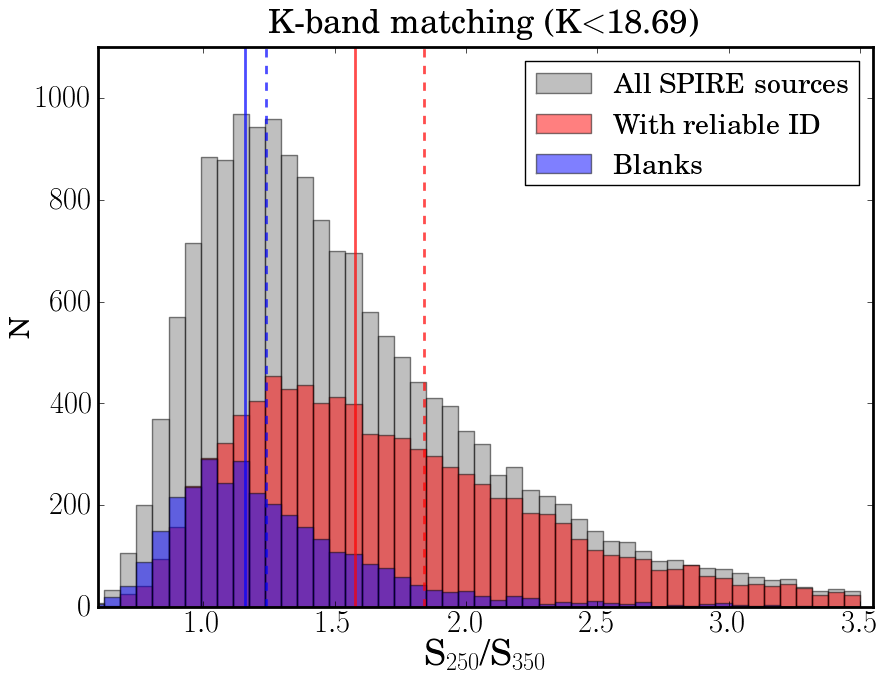}
\caption{$S_{\rm 250}/S_{\rm 350}$ colour distribution of of the SPIRE sources. The grey histogram corresponds to the colour distribution of all SPIRE sources with ${\rm SNR}_{250}\ge4$. The red histogram corresponds to those sources with a $K<18.69$ reliable match and the blue histogram to those with no match (blanks).For comparison, the vertical lines correspond to the median values of the distributions from Fig. \ref{colour_dist_rel_unrel}. }
\label{col_dist_rel_unrel_newdepth}
\end{figure}
\begin{figure}
\includegraphics[scale=0.35]{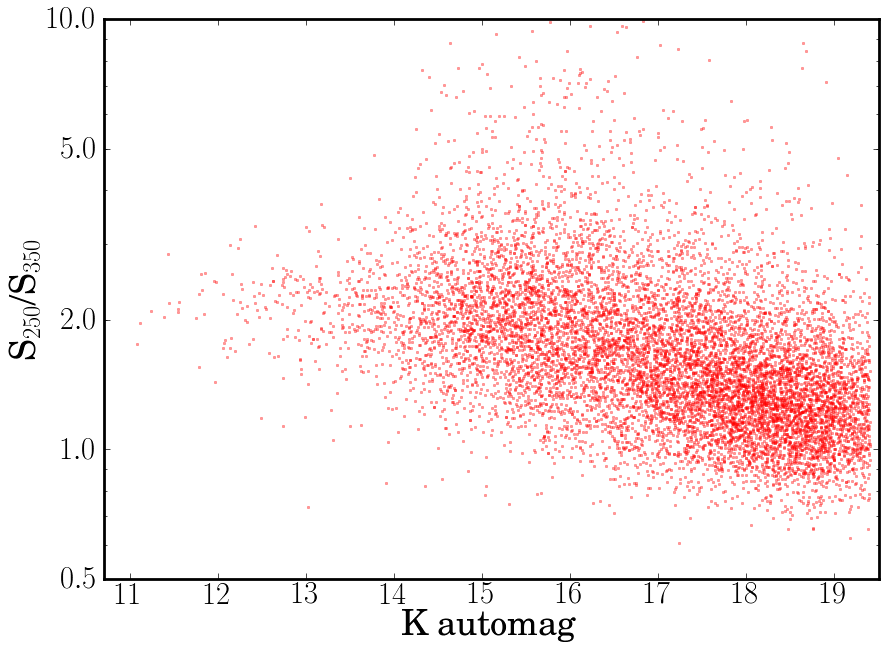}
\caption{ $S_{\rm 250}/S_{\rm 350}$ colour of the SPIRE sources reliably identified in the deeper $K$-band catalogue ($K<19.40$) as a function of the counterpart $K$ magnitude. Fainter counterparts are associated to redder SPIRE sources.  }
\label{mag_col_diagram_k}
\end{figure}

In Fig. \ref{col_dist_rel_unrel_newdepth} we present the colour distribution of SPIRE sources with reliable counterparts identified in the shallower $K$-band catalogue and of the blank fields. This is similar to the lower panel of Fig. \ref{colour_dist_rel_unrel}, but now considering the NIR identifications with $K<18.69$ to SPIRE sources, resulting from the new LR analysis. For comparison, we plotted the median values of the colour distributions of Fig. \ref{colour_dist_rel_unrel} as vertical lines. The SPIRE sources reliably identified in the shallower $K$-band catalogue are distinctively redder than the ones identified in the optical catalogue (upper panel of Fig. \ref{colour_dist_rel_unrel}). Comparing to the lower panel of Fig. \ref{colour_dist_rel_unrel}, it is easy to see that by using a shallower $K$-band catalogue we are missing more red SPIRE sources than blue ones. This becomes more evident in Fig. \ref{mag_col_diagram_k}, where we show the colour of the SPIRE sources reliably identified in the deeper $K$-band catalogue versus the magnitude of the reliable counterpart. Fainter near-infrared reliable counterparts are associated with redder SPIRE sources. 

\begin{figure*}
\includegraphics[scale=0.35]{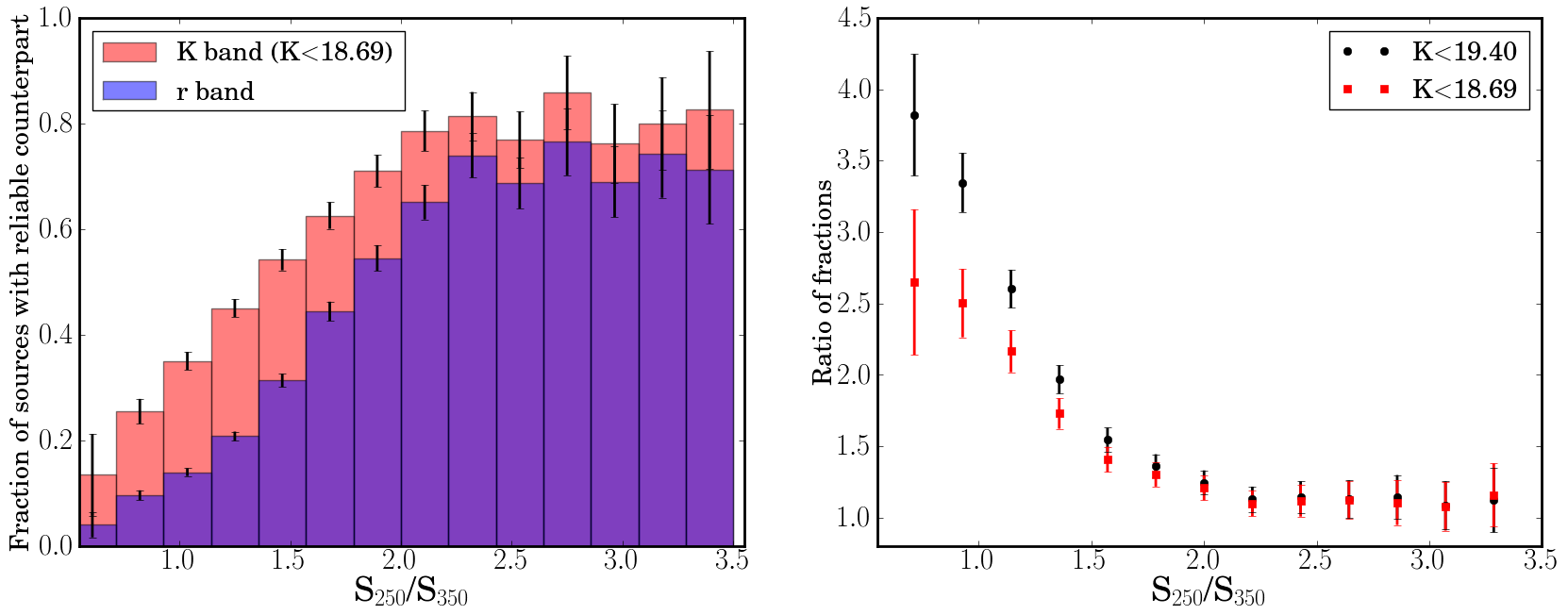}
\caption{{\it Left panel:} Fraction of SPIRE sources with ${\rm SNR}_{250}\ge4$ with a reliable counterpart identified in the optical and shallower near-infrared ($K<18.69$) catalogues as a function of the source $S_{\rm 250}/S_{\rm 350}$ colour. {\it Right panel:} Ratio between the near-infrared and optical fractions. The red squares correspond to the ratio of the fractions shown in the left panel. The black circles correspond to the case where the reliable identifications in the deeper $K$ catalogue ($K<19.40$) are considered. The error bars represent Poisson counting errors.}
\label{fraction_sources_rel_count}
\end{figure*}

We then compare the fraction of SPIRE sources with a reliable counterpart identified in the optical and NIR shallower object catalogues (with matched source density in the $25.93$ deg$^2$ overlapping area) as a function of the $S_{\rm 250}/S_{\rm 350}$ colour in the left panel of Fig. \ref{fraction_sources_rel_count}. For bluer SPIRE sources ($S_{\rm 250}/S_{\rm 350} >2.4$), the fraction of sources reliably identified is above $70$ per cent for both $r$- and $K$-band matchings. The fraction of SPIRE sources with a reliable identification decreases towards redder colours. However, this decrease is much more significant for the optical identifications than for the near-infrared ones. For redder SPIRE sources ($S_{\rm 250}/S_{\rm 350}<1.2$), the fraction of SPIRE sources reliably identified in the shallower $K$ catalogue ($K<18.69$) is more than 2.5 times higher than the fraction of SPIRE sources reliably identified in SDSS, as shown in the right panel of the figure (red squares), where the ratio between the near-infrared and optical reliable fractions is shown. Considering the reliable identifications in the deeper $K$ catalogue ($K<19.40$), this ratio can be higher than 3.5 (black circles). 

\begin{figure}
\includegraphics[scale=0.35]{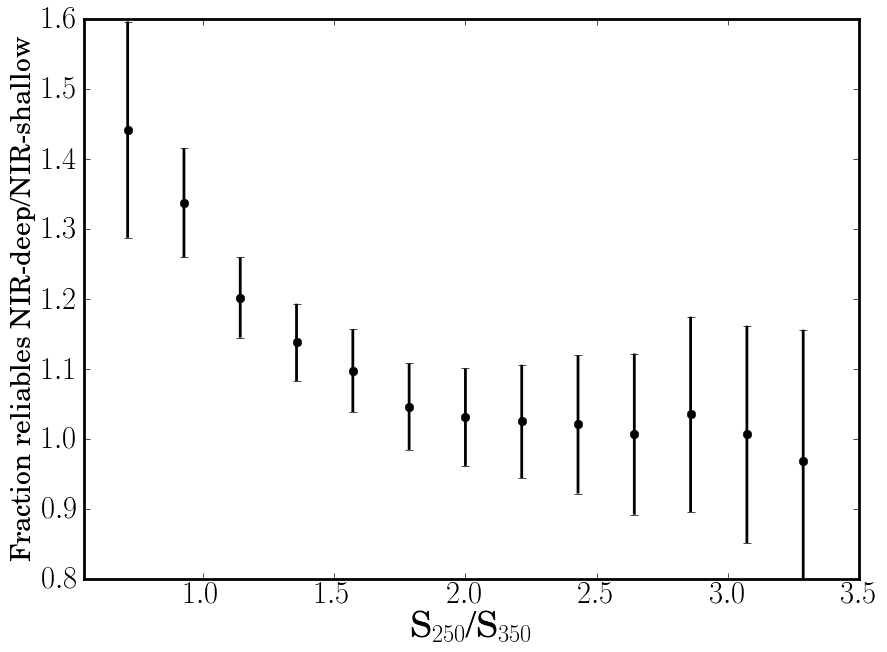}
\caption{Ratio between the fraction of SPIRE sources with a reliable counterpart identified by the LR method in the deeper ($K<19.40$) catalogue and the fraction of SPIRE sources with a reliable counterpart identified in shallower ($K<18.69$) catalogue as a function of the SPIRE $S_{\rm 250}/S_{\rm 350}$ colour. The error bars correspond to the Poisson errors on the ratios.}
\label{fraction_rel_colour_k_shallow_deep}
\end{figure}

In order to quantify the effect of the increasing of the depth of the input catalogue on the identification of the red SPIRE sources we compared the fraction of SPIRE sources with a reliable counterpart identified applying the LR method to shallower ($K<18.69$) and deeper ($K<19.40$) $K$-band catalogues. In Fig. \ref{fraction_rel_colour_k_shallow_deep} we show the ratio of those fractions. For the bluer SPIRE sources, a similar fraction of reliable identifications is found in both the shallower and deeper $K$-band catalogue. The bluer SPIRE sources are generally associated with bright NIR objects and increasing the depth of the NIR catalogue does not improve the detection of those sources. For the redder SPIRE sources, the depth of the catalogue has a important role on the identification of their NIR counterparts. Going 0.7 mag deeper increases the fraction of redder SPIRE sources ($S_{\rm 250}/S_{\rm 350}<1.2$) reliably identified by more than $20$ per cent. The redder SPIRE sources are likely to be at higher redshifts and can be too faint to be detected in $K<18.69$ catalogue. As discussed in \citet{bourne14}, the larger positional error of red sources is due to lensing (the red submm sources being falsely identified to a low-redshift lensing structure) and/or clustering effects (red sources constitute a more clustered population than the blue ones). We found that this effect was weaker in the NIR than in the optical. The larger $\sigma_{\rm pos}$ values for red SPIRE sources when fitting the SPIRE-optical offsets (see Fig. \ref{sigma_pos_cbins}) can be explained by the lack of some true counterpart in the optical, inducing the cross-correlation between the foreground lensing structures and background high-$z$ SPIRE sources \citep{gonzalez17}. 

\begin{figure}
\includegraphics[scale=0.35]{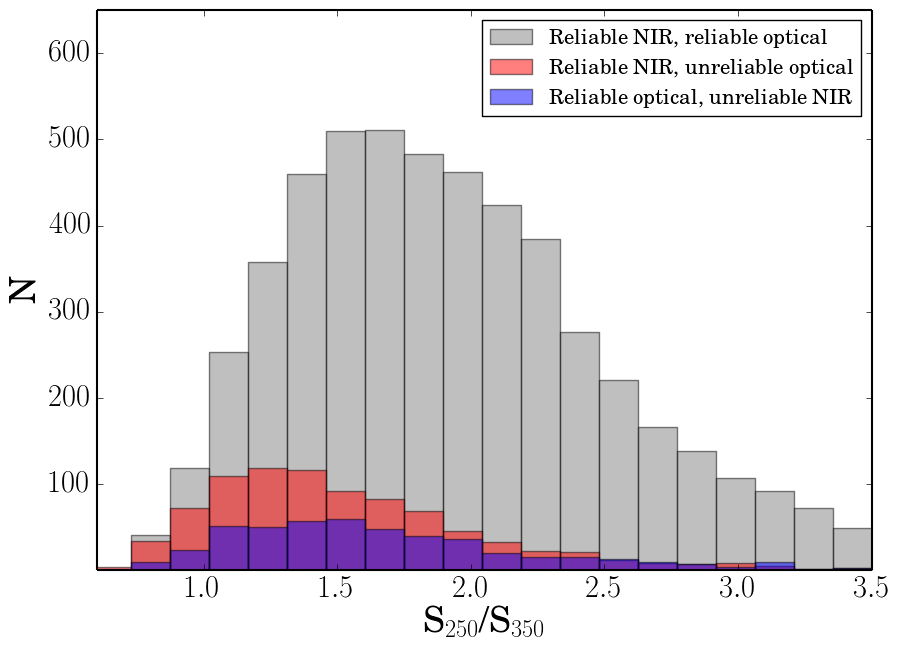}
\caption{ $S_{\rm 250}/S_{\rm 350}$ colour of the SPIRE sources whose reliabilities in the optical and NIR agree or disagree. The grey histogram shows the colour distribution of the SPIRE sources reliably identified in both wavelengths. The red histogram shows the colour distribution of the SPIRE sources identified as reliable in the NIR but as unreliable in the optical. The colour distribution of the opposite situation (SPIRE sources identified as reliable in the optical but as unreliable in the NIR) is shown in blue.  }
\label{colour_distribution_NIR_optical}
\end{figure}

Finally, it is interesting to quantify how many SPIRE sources the optical and NIR matching agree or disagree on reliable identifications. In order to compare our results, we cross-matched our SDSS candidate IDs with the shallower $K$-band ID catalogue. We used a 2 arcsec search radius and considered the nearest match. We checked that for all the matches, both optical and NIR counterpart are associated with the same SPIRE source. We found that 9388 of the 23341 NIR counterparts are not detected in the optical ID catalogue within 2 arcsec. Conversely, 3154 of 17107 optical identifications are not detected in the NIR ID catalogue within the same radius search. There are 5497 SPIRE sources reliably identified in both ID catalogues. This corresponds to $\sim$ 88 per cent of the reliable optical matches and to $\sim$ 58 per cent of the $K$-band reliable matches. 503 reliable identifications in the optical are classified as unreliable in the NIR ID (although 234 of them have $R\ge 0.6$), while 902 reliable NIR candidates are classified as unreliable in the optical ID catalogue (539 of them have $R \ge 0.6$). We found that 3060 SPIRE sources reliably identified in the NIR are blank in the SDSS. A much smaller number (285) of SPIRE sources identified in the optical are blank in the NIR catalogue. In Fig. \ref{colour_distribution_NIR_optical} we show the colour distribution of SPIRE sources whose reliabilities in the optical and NIR agree or disagree. The subset of SPIRE sources identified as reliable in the NIR but as unreliable in the optical is slightly redder (median $S_{\rm 250}/S_{\rm 350}=$ 1.45) than the opposite case, where SPIRE sources identified as reliable in the optical but as unreliable in the NIR (median $S_{\rm 250}/S_{\rm 350}=1.59$). The fact that both subsets (where a reliable ID was only found in either optical or NIR) have redder colours than the subset of IDs which are reliable in both catalogues indicates that its more likely that the identification is ambiguous or wrong if the SPIRE colour is redder.

The near-infrared bands offer a significant improvement for the identification of counterparts to SPIRE sources, particularly the red ones. 

\section{Radio counterparts to submm sources}
\label{sec_id_radio}

Radio observations are particularly useful to unambiguously identify counterparts to submm sources. The low surface density of radio sources combined with the significantly higher resolution offered via interferometry reduces the probability of a chance association to a submm source. Moreover, searching for the radio emission of submm sources exploits the well known correlation between radio and far-infrared (FIR) emission \citep{helou85,condon91,ivison10}. Once a radio counterpart of a submm source is identified, its accurate position can be used to easily identify the source at other wavelengths. The disadvantage of matching to radio counterparts is that, because of the radio-FIR correlation, radio counterparts are more likely to be found for the brightest submm sources, which may not represent the entire population. Another disadvantage is that a large fraction of submm sources do not have radio counterparts \citep{chapman03,barger07,Ivison2007,Ibar10a}. Because the radio observations do not benefit from a negative K-correction, normal galaxies (with $\alpha=0.7$, where $S\propto\nu^{-\alpha}$; e.g. \citealt{Ibar09}) at high redshift ($z>3$) are very difficult to detect at 1.4 GHz. This means that submm sources without a radio counterpart are likely to be at higher $z$. 

Unfortunately, using radio interferometric data as a gateway to identifying optical and NIR counterparts in large area submm surveys is not yet practical with the current instruments (e.g. FIRST, SUMMS, NVSS, etc). Upcoming and future radio surveys, such as the Square Kilometre Array (SKA) and the Low Frequency Array (LOFAR) will be able to provide the required depth and area, so that this method can be extended to larger areas. 

The radio data were obtained during nine transits of the Subaru Deep Field (SDF) in 2009 March and April using National Radio Astronomy Observatory Very Large Array (VLA), in its B configuration (Project AM979).  Data were recorded every 3.3\,sec, with $14\times 3.125$-MHz dual-polarisation channels, centred roughly at 1,315\,MHz, or 22.8\,cm.  Of 42\,hr in total spent on-sky, around 36\,hr comprised 30-min scans of a single pointing in SDF ($13^{\rm h} 24^{\rm m} 38.^{\rm s}9, +27^\circ 29' 25".9$ J2000), sandwiched between 2-min scans of 3C\,286, which was used to calibrate the phase and amplitude of the visibilities, as well as the bandpass, and to set the absolute flux density scale. 

The flagging, calibration and imaging of these data followed the procedures outlined in \citet{biggs2006}, yielding a map covering the VLA's primary beam (31', FWHM), with 1-arcsec$^2$ pixels.  No correction was applied for the primary beam response.  The r.m.s.\ noise level in regions close to the pointing centre, free of sources, is around 11\,$\mu$Jy\,beam$^{-1}$.  A catalogue of 602 sources with peak flux densities above $5\,\sigma$ was then selected from a field of radius 24\,arcmin.

We start by identifying the radio counterparts to the SPIRE sources. We used the $p$ statistics \citep{downes86} to quantify the likelihood of a radio source being a chance association to the SPIRE source. This procedure is more appropriate to identify radio counterparts than the LR method, since the surface density of radio sources is much lower than the optical and NIR catalogues. The $p$-values are calculated as
\begin{equation}
p=1-exp(-\pi n \theta^2)
\label{pstat}
\end{equation}
where $n$ is the surface density of radio sources and $\theta$ is the radio-SPIRE positional offset. We searched for radio counterparts within 10 arcsec from the SPIRE sources. We define robustly matched radio sources as those satisfying $p<0.05$. 

\begin{figure*}
\includegraphics[scale=0.35]{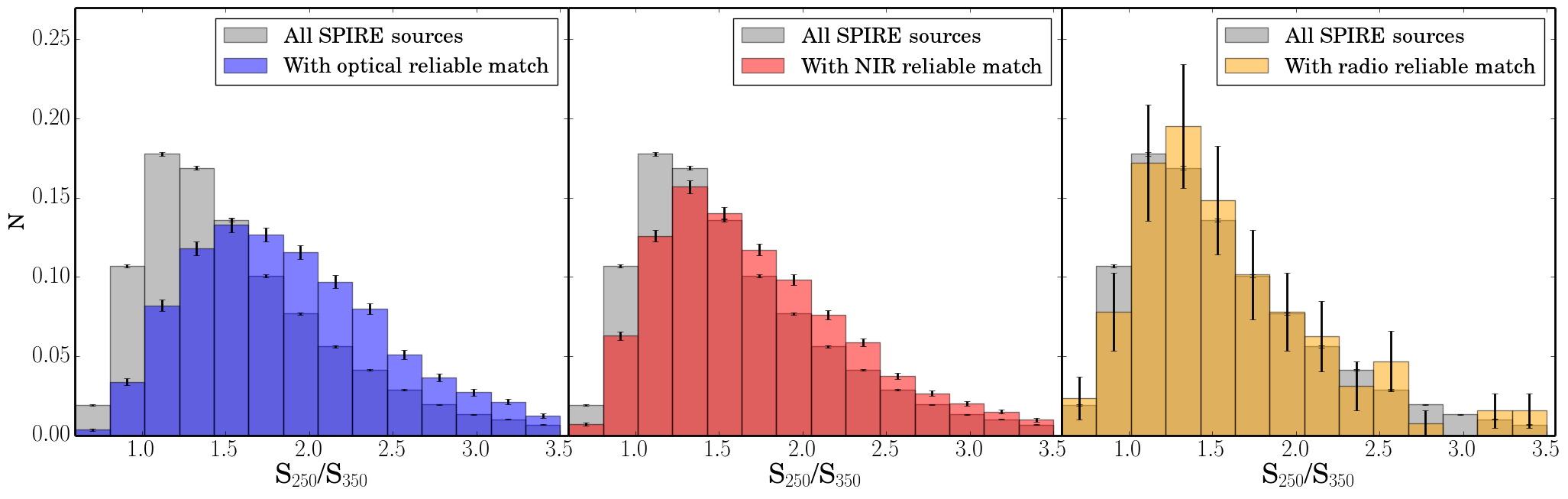}
\caption{Normalized colour distribution of the SPIRE sources with optical (left panel), NIR (central panel) and radio (right panel) reliable counterparts identified. }
\label{colour_dist_opt_nir_radio}
\end{figure*}

Using this technique we found that 128 SPIRE sources with SNR$_{250}>4$ have at least one robust radio counterpart. This corresponds to $\sim$ 45 per cent of SPIRE sources  with SNR$_{250}>4$ in the SDF area. We found 139 robust radio identifications to those 128 SPIRE sources. There are 11 SPIRE sources with 2 robust radio counterparts. The colour distribution of SPIRE sources with radio counterparts follows the overall colour distribution of the SPIRE sources, as shown in Fig. \ref{colour_dist_opt_nir_radio}. In this figure, we also show the normalized colour distribution of SPIRE sources with optical and NIR reliable identifications. A Kolmogorov-Smirnov (K-S) test was used to determine whether the colour distribution of SPIRE sources with radio counterparts differ significantly from the overall distribution. The result indicates that the two distributions are statistically indistinguishable. 

We then matched the position of our robust radio identifications to the positions of the optical and NIR IDs. For a fair comparison, we used the NIR ID catalogue resulting from the LR analysis based on the shallower NIR data, with a source density matching the SDSS catalogue. We found that 84 SPIRE sources with a robust radio counterpart also have a reliable ID in the NIR. Of those, 73 (87 per cent) lie within 2 arcsec of the radio source. For the radio-optical matching, we found that 51 SPIRE sources with a robust radio counterpart also have a reliable ID in the optical, of which 42 (82 per cent) have offset smaller than 2 arcsec. Assuming that the radio identification is the true counterpart, these results indicate that more than the $80$ per cent of the reliable identifications provided by the likelihood method are correct. This estimate of the cleanness of the reliable IDs from the likelihood method is smaller than the one obtained from equation (\ref{cleanness}), $\sim95$ per cent for both optical and NIR cases. 
A larger radio data set is required to confirm these results. 

\begin{figure}
\includegraphics[scale=0.35]{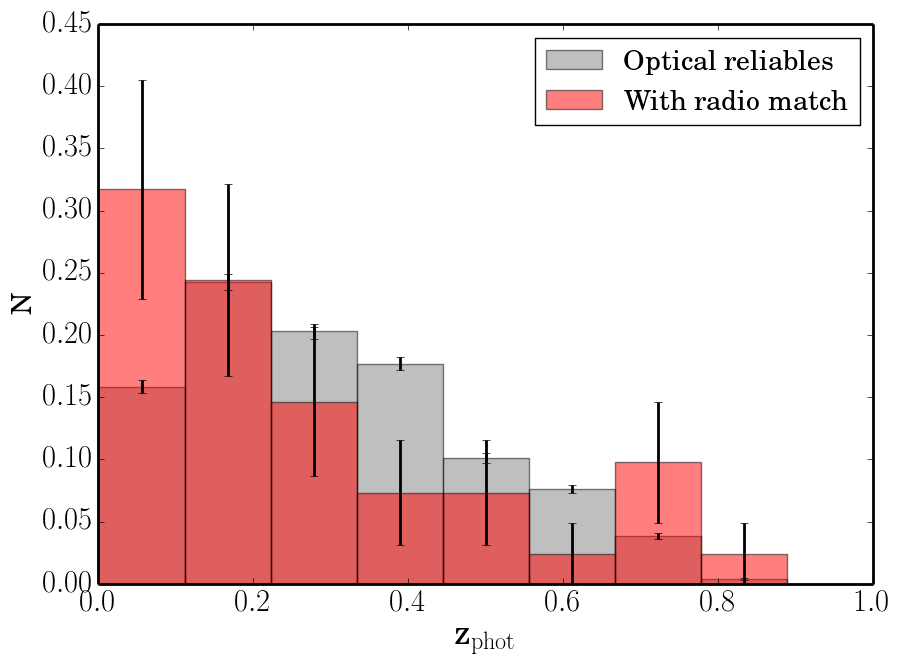}
\caption{Normalized photometric redshift distribution of all optical reliable counterparts (grey histogram) and of those that match to a robust radio identification (red histogram). }
\label{photoz_dist_optrel_radio}
\end{figure}

We also investigated the photometric redshifts of the robust radio identifications. In Fig. \ref{photoz_dist_optrel_radio} we compare the photometric redshift distribution of all optical reliable counterparts to that of the optical reliable counterparts that also match to a robust radio identification. 

The results from Fig. \ref{colour_dist_opt_nir_radio} indicate that the radio is less biased against identifying redder SPIRE sources than the NIR and optical. This is expected given the correlation between the radio and far-infrared emission. Our radio catalogue is deep enough to detect SPIRE sources at higher redshifts, as it can be seen in Fig. \ref{photoz_dist_optrel_radio}.

\section{Conclusions}
\label{sec_conclusions}

In this paper we described the results of the identification of multi-wavelength counterparts to submm sources in the NGP area. 

We described the process to obtain the SDSS counterparts at $r<22.4$ for the $250$-$\mu$m-selected sources detected at $\ge 4 \sigma$ over the entire NGP field. We used the likelihood ratio method to measure the reliability of all potential counterparts within $10$ arcsec of the SPIRE sources. We obtained, using a blank-field comparison, that the fraction of SPIRE sources that should have a counterpart detection with $r<22.4$ in SDSS is $Q_0=0.538 \pm 0.001$, which is in agreement with the results obtained by \citetalias{bourne16} for the GAMA fields in the Phase 1 H-ATLAS data release. We determined reliable ($R\ge 0.8$) optical counterparts to 42429 SPIRE sources, corresponding to 37.8 per cent of the SPIRE sources with SNR$\ge 4$ in the NGP field. We estimated that this sample has a completeness of 70.3 per cent and a false identification rate of 4.7 per cent.

We also identified counterparts to the $250$-$\mu$m-selected sources in the near-infrared using $K$-band data obtained with UKIRT/WFCAM in a smaller $\sim 25$ deg$^2$ area within NGP using the likelihood ratio method. We searched for near-infrared counterparts with $K<19.4$ within $10$ arcsec of the SPIRE sources with SNR$\ge 4$. We estimated that the fraction of all counterparts which are above the magnitude limit ($K<19.4$) is $Q_0=0.836 \pm 0.001$ ($0.0754\pm 0.001$ for extragalactic objects and $0.08\pm0.001$ for stars). We were able to find a reliable match to 61.8 per cent of the SPIRE sources within that area. We estimated an overall completeness of 74 per cent and cleanness of $95.5$ per cent for this near-infrared reliable sample.

We visually inspected the brightest sources in the our optical and near-infrared ID catalogues in order to validate the automated counterpart identification process and to correct missing or misclassified ID information when necessary. 

We investigated the performance of the likelihood ratio method to identify optical and near-infrared counterparts. Using input catalogue for the matching with the same surface density, we compared the optical and near-infrared counterparts identified by the likelihood ratio technique in the region of NGP with SDSS and WFCAM coverage. We obtained that 54.8 per cent of the $250$-$\mu$m selected sample have an near-infrared reliable counterpart with $K<18.69$. This reliable fraction is much higher than the one obtained for the optical matching, in which we were able to reliably match 36.4 per cent of the SPIRE sources in the overlapping area. This result confirms that near-infrared bands are much better placed than optical bands to reliably identify submm sources, although some reliable counterparts found in the optical are not found in the K band. Moreover, $K$-band is particularly useful to identify redder SPIRE sources. The fraction of redder ($S_{\rm 250}/S_{\rm 350}<1.2$) SPIRE sources reliably identified in the shallower $K$ catalogue ($K<18.69$) is more than 2.5 times higher than the fraction of SPIRE sources reliably identified in SDSS. We demonstrated that fainter near-infrared reliable counterparts are generally associated to redder SPIRE sources and that going $0.7$ deeper in the $K$ magnitude limit (from $K=18.69$ to $K=19.40$) the fraction of redder SPIRE sources reliably identified is increased by $\sim 20-45$ per cent.  

Finally, we assessed the efficacy of our identification of optical and near-infrared counterparts by using deep radio interferometric data. We used $p$-values to find the radio counterparts to the SPIRE sources. We determined robust ($p<0.05$) radio identifications to 128 SPIRE sources. We matched the position of the robust radio identifications to the positions of the optical and NIR IDs. Assuming the radio counterpart as the true one, we obtained that $80-90$ per cent of the reliable identifications provided by the likelihood method for both optical and near-infrared matching are correct. Future large area radio surveys will be able to confirm this result for a larger sample of submm sources.

\section*{Acknowledgements}

The authors thank Soh Ikarashi for the providing radio maps and catalogues and Mike Read for stacking 
the UKIRT/WFCAM data used in this work.
SD is supported by an STFC Ernest Rutherford Fellowship. 
EV and SAE acknowledge funding from the UK Science and Technology Facilities Council consolidated grant ST/K000926/1.
MS and SAE have received funding from the European Union Seventh Framework Programme ([FP7/2007-2013] [FP7/2007-2011]) under grant agreement No.~607254.
SM and LD acknowledge support from the European Research Council (ERC) in the form of the Consolidator Grant {\sc CosmicDust} (ERC-2014-CoG-647939, PI H.L.Gomez).
NB, SM, LD and RJI acknowledge support from the ERC in the form of the Advanced Investigator Program, {\sc COSMICISM} (ERC-2012-ADG\_20120216, PI R.J.Ivison). EI\ acknowledges partial support from FONDECYT through grant N$^\circ$\,1171710. {\it Herschel} is
an ESA space observatory with science instruments provided by
European-led Principal Investigator consortia and with important
participation from NASA. Funding for SDSS-III has been provided by the
Alfred P. Sloan Foundation, the Participating Institutions, the
National Science Foundation, and the U.S. Department of Energy Office
of Science. The SDSS-III web site is http://www.sdss3.org/. SDSS-III
is managed by the Astrophysical Research Consortium for the
Participating Institutions of the SDSS-III Collaboration including the
University of Arizona, the Brazilian Participation Group, Brookhaven
National Laboratory, Carnegie Mellon University, University of
Florida, the French Participation Group, the German Participation
Group, Harvard University, the Instituto de Astrofisica de Canarias,
the Michigan State/Notre Dame/JINA Participation Group, Johns Hopkins
University, Lawrence Berkeley National Laboratory, Max Planck
Institute for Astrophysics, Max Planck Institute for Extraterrestrial
Physics, New Mexico State University, New York University, Ohio State
University, Pennsylvania State University, University of Portsmouth,
Princeton University, the Spanish Participation Group, University of
Tokyo, University of Utah, Vanderbilt University, University of
Virginia, University of Washington, and Yale University.


\label{lastpage}

\end{document}